\documentclass[twocolumn, prl, superscriptaddress,notitlepage]{revtex4-2}
\usepackage{graphicx}
\usepackage{dcolumn}
\usepackage{bm}
\usepackage[usenames,dvipsnames]{color}
\usepackage[most]{tcolorbox}
\usepackage{multirow}
\usepackage{gensymb}
\usepackage[normalem]{ulem}
\usepackage{CJK}
\usepackage{comment}
\usepackage[colorlinks, linkcolor=blue,anchorcolor=blue,citecolor=blue,urlcolor=blue]{hyperref}
\usepackage{amssymb}
\usepackage{pifont}
\usepackage{upgreek}
\usepackage{physics}
\usepackage{natbib}
\usepackage{xcolor}

\usepackage{color,soul}
\usepackage{tikz}
\usetikzlibrary{decorations.markings}
\usetikzlibrary{arrows.meta,calc,positioning,shapes.geometric}

\begin{document}
\begin{CJK*}{UTF8}{}
\title{Spin Interferometry Generated by Cavity Memory}
\author{Siyu Li}\thanks{These authors contributed equally.}
\affiliation{International Center for Quantum Materials, School of Physics, Peking University, Beijing 100871, China}
\affiliation{Beijing Key Laboratory of Quantum Devices, Peking University, Beijing 100871, China}
\author{Yifan Shi}\thanks{These authors contributed equally.}
\affiliation{International Center for Quantum Materials, School of Physics, Peking University, Beijing 100871, China}
\affiliation{Beijing Key Laboratory of Quantum Devices, Peking University, Beijing 100871, China}

\author{Yicheng Xiao}
\affiliation{International Center for Quantum Materials, School of Physics, Peking University, Beijing 100871, China}
\affiliation{Beijing Key Laboratory of Quantum Devices, Peking University, Beijing 100871, China}

\author{Haotian Xu}
\affiliation{International Center for Quantum Materials, School of Physics, Peking University, Beijing 100871, China}
\affiliation{Beijing Key Laboratory of Quantum Devices, Peking University, Beijing 100871, China}

\author{Shixuan Zhang}
\affiliation{International Center for Quantum Materials, School of Physics, Peking University, Beijing 100871, China}
\affiliation{Beijing Key Laboratory of Quantum Devices, Peking University, Beijing 100871, China}

\author{Aoqin Liang}
\affiliation{International Center for Quantum Materials, School of Physics, Peking University, Beijing 100871, China}
\affiliation{Beijing Key Laboratory of Quantum Devices, Peking University, Beijing 100871, China}

\author{Yanxi Wu}
\affiliation{International Center for Quantum Materials, School of Physics, Peking University, Beijing 100871, China}
\affiliation{Beijing Key Laboratory of Quantum Devices, Peking University, Beijing 100871, China}

\author{Xiaokang Li}
\affiliation{International Center for Quantum Materials, School of Physics, Peking University, Beijing 100871, China}
\affiliation{Beijing Key Laboratory of Quantum Devices, Peking University, Beijing 100871, China}

\author{Guoqing Wang} \email[]{gq\_wang@pku.edu.cn}
\affiliation{International Center for Quantum Materials, School of Physics, Peking University, Beijing 100871, China}
\affiliation{Beijing Key Laboratory of Quantum Devices, Peking University, Beijing 100871, China}

\begin{abstract}
We report the control and observation of coherent spin interferometry generated by cavity memory using an ensemble of nitrogen-vacancy (NV) centers coupled to a microwave cavity. The abrupt switch-on and switch-off of a detuned drive produce transient fields at the cavity frequency that supply the two spin rotations for Ramsey interferometry, observed using both microwave and fluorescence readout. Spectrally scanning the spin profile across the cavity and drive frequencies distinguishes cavity-generated Ramsey interference from conventional Rabi dynamics. Temporal weighting of the reflected microwave signal enables spectrally selective population readout at a single probe frequency. By programming the phase and amplitude of the drive envelope, we extend the same mechanism to more quantum control sequences, including spin-echo and Rabi control. 
\end{abstract}

\maketitle

\end{CJK*}

\textit{Introduction.---} Cavity-coupled spin ensembles have emerged as a versatile quantum platform, whose collective strong coupling~\cite{kubo_strong_2010,schuster_highcooperativity_2010,amsuss_cavity_2011, probst_anisotropic_2013,ranjan_probing_2013} has enabled applications such as microwave
storage and coherence protection~\cite{zhu_coherent_2011,grezes_multimode_2014,putz_protecting_2014}. The same collective interface has revealed rich non-equilibrium phenomena, including superradiance and subradiance, self-stimulated echo trains, self-induced and sustained superradiant masing, as well as dissipative many-body dynamics~\cite{angerer_superradiant_2018,weichselbaumer_echo_2020,debnath_selfstimulated_2020,kersten_triggered_2023,kersten_selfinduced_2026,wu_sustained_2026,lei_manybody_2023}. In parallel, quantum-limited, dispersive and cavity-enhanced microwave readout has enabled sensitive detection of spin ensembles, including room-temperature NV magnetometry with near-unity microwave contrast and sub-picotesla sensitivity~\cite{bienfait_reaching_2016,ebel_dispersive_2021,eisenach_cavityenhanced_2021,wang_spinrefrigerated_2024}. Maser-based spin detection~\cite{wu_enhanced_2022,han_ultrasensitive_2026} and nonlinear spin--oscillator dynamics near a bistable transition~\cite{wang_exceptional_2026} have further extended the sensing capabilities of these hybrid systems. These advances establish cavity-coupled spin ensembles as an important platform for quantum memories, sensing and simulation.

Despite this progress, phase coherent pulsed control in a narrow-band cavity remains challenging. Although resonator-enhanced Rabi control has been demonstrated using dielectric resonators with relatively broad linewidths~\cite{kapitanova_3d_2018,vallabhapurapu_fast_2021,alsid_solidstate_2023}, reducing cavity loss to increase spin-cavity cooperativity for strong coupling motivates the use of higher-quality-factor cavities~\cite{kubo_strong_2010,schuster_highcooperativity_2010}. However, the resulting narrower cavity linewidth and the corresponding longer cavity lifetime prevent the intracavity field from instantaneously following an external waveform. Each change of the applied drive leaves a history-dependent field in the cavity, reshaping the amplitude and phase of control pulses and distorting subsequent spin rotations. Pulse shaping and optimal control approaches account for resonator distortions, suppress ringdown, and address the dynamics of inhomogeneous spin ensembles coupled to a cavity~\cite{borneman_bandwidthlimited_2012,probst_shaped_2019,ansel_optimal_2018,iriarte-zendoia_robust_2025}. From a different point of view, cavity transients and pulse-edge effects have been exploited as useful resources for interferometric optical sensing with cavity ring-up spectroscopy~\cite{rosenblum_cavity_2015}, resonator characterization~\cite{gyure_time_2015}, and the generation and subsequent echo retrieval of spin coherences in strongly coupled molecular spin--cavity systems~\cite{lenz2021room}. 


In this Letter, we show that the finite cavity response to a detuned drive can directly supply coherent rotations needed for spin interferometry. 
Using an ensemble of NV centers coupled to a room-temperature microwave cavity, we realize Ramsey interferometry by applying a detuned microwave drive pulse. The abrupt switching at the two edges of the drive pulse generates transient fields at the cavity frequency, which provide the preparation and analysis spin rotations for Ramsey interferometry. The observed fringes oscillate at a frequency set by the drive-cavity detuning. Spectral scans using independent fluorescence and microwave readout distinguish the Ramsey interference from conventional Rabi dynamics induced directly by the applied drive. In particular, by analyzing the temporal dependence of cavity reflection, we extract signals from different spectral components of the spin profile and realize spectrally selective microwave readout. Furthermore, we extend this approach to more quantum control sequences by engineering the phase and amplitude of the drive envelope, including a spin echo implemented via a drive-phase jump and Rabi oscillations produced by suppressing the switch-off transient via an adiabatic ramp-down. Our work provides a new route to pulsed spin interferometry and programmable quantum control in cavity-coupled hybrid quantum systems.

\begin{figure*}[htbp]
    \centering
    \includegraphics[width=\textwidth]{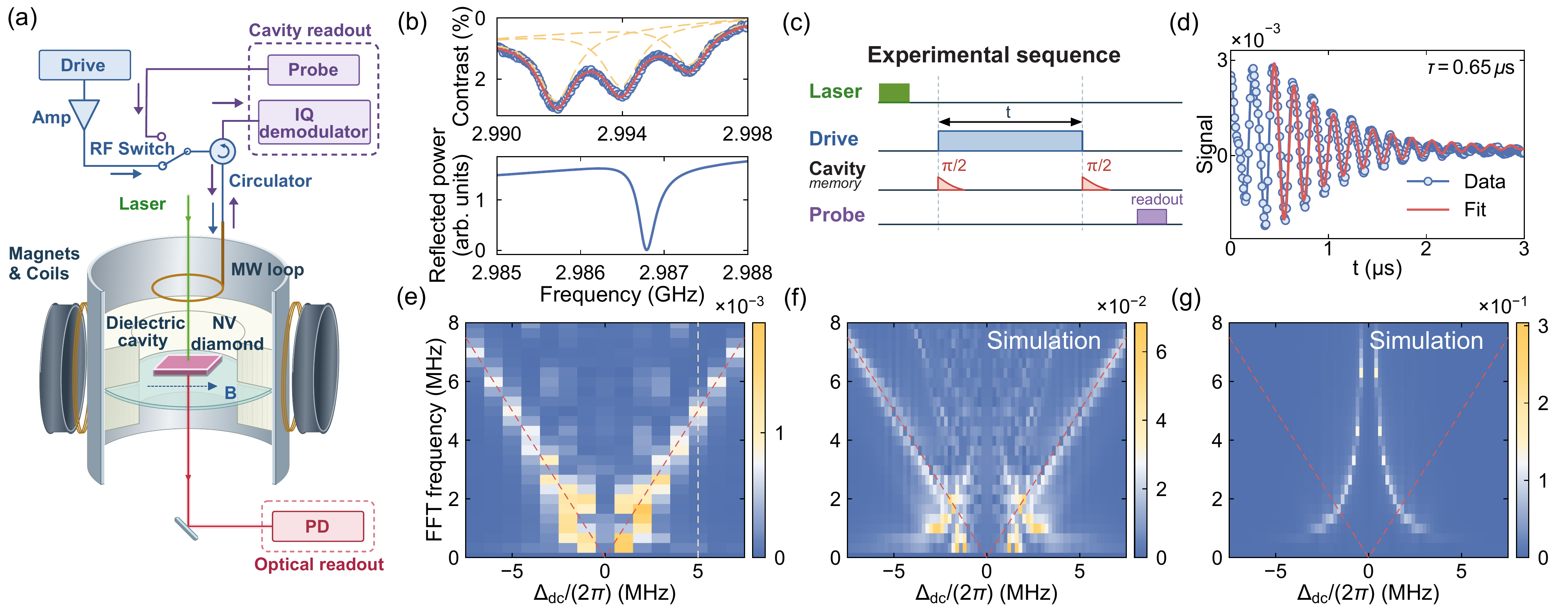}
    \caption{\textbf{Ramsey interference generated by cavity memory.} 
    (a) Schematic of the experimental setup. 
    (b) Top: optically detected magnetic resonance spectrum and corresponding fits to the hyperfine components. Bottom: cavity reflection spectrum. 
    (c) Experimental sequence for Ramsey interferometry generated by cavity memory. The two switching-induced cavity transients act as effective $\pi/2$ rotations separated by an evolution time $t$.
    (d) A representative Ramsey interference signal.  Drive parameters are $\Omega_{\mathrm{d}}/(2\pi)=0.53\,\mathrm{MHz}$ and $\Delta_{\mathrm{dc}}/(2\pi)=5 \,\mathrm{MHz}$. The central hyperfine transition ($m_I=0$) is tuned to the cavity resonance. The red curve fits the experimental data to an exponentially decaying oscillation model with a decay time $\tau=0.65\,\upmu \mathrm{s}$. 
    (e) Measured Fourier spectra of the Ramsey signals versus $\Delta_{\mathrm{dc}}$. The applied drive power is fixed, which is characterized by an on-cavity-resonance Rabi frequency $\Omega_0/(2\pi)=21.4\,\mathrm{MHz}$, and the intracavity drive amplitude is detuning dependent. The spectrum at the white dashed line corresponds to the data shown in (d). The experimental signals are obtained from the quadrature channel of the IQ demodulator under a cavity-resonant probe. (f) Simulated Fourier spectra of time evolutions of the total spin population in $m_S=0$ under the same experimental conditions in (e), taking into account the inhomogeneity of the spin ensemble. 
    (g) Simulated Fourier spectra of time evolutions of the spin population without including the effects of cavity transients, while keeping the other parameters the same as (f). }
    \label{fig:1}
    
\end{figure*}

\textit{Experimental setup.---} We couple an inhomogeneously broadened NV ensemble to a dielectric microwave cavity [Fig.~\ref{fig:1}(a)]. A pair of permanent magnets provides a static magnetic field of $68\,\mathrm{G}$ along the diamond $[100]$ axis, making the four NV orientation groups spectrally degenerate. The cavity supports a $\mathrm{TE}_{01\delta}$ mode with a longitudinal microwave magnetic field at $\omega_c/(2\pi)=2.9868\,\mathrm{GHz}$. Its full width at half maximum linewidth is $\kappa/(2\pi)=0.25\,\mathrm{MHz}$, corresponding to a loaded quality factor $Q_L \simeq1.2\times10^4$. A pair of coils fine-tunes the spin-cavity detuning. For optical initialization, a $532\,\mathrm{nm}$ laser beam delivers $5\,\mathrm{W}$ over an illuminated area of approximately $1\,\mathrm{mm}^2$. The ensemble can be measured through either fluorescence collected by a photodiode or the cavity reflection of a weak microwave probe, whose amplitude and phase are detected with an IQ demodulator (see supplement for details~\cite{SOM}). The optically detected magnetic resonance spectrum resolves the hyperfine structure and yields an inhomogeneous linewidth of $\gamma/(2\pi)=1.385\,\mathrm{MHz}$ [Fig.~\ref{fig:1}(b), top panel]. Three dips correspond to three hyperfine transitions with distance $A_{zz}/(2\pi) = 2.16$ MHz. The cavity reflection spectrum is shown by the bottom panel of Fig.~\ref{fig:1}(b). 
An amplified microwave drive provides coherent control, while a switch routes either the drive or probe channel to the same coupling loop.

\textit{Ramsey interference generated by cavity memory.---}
We observe Ramsey interference when applying a single-pulse sequence conventionally used to measure Rabi oscillations [Fig.~\ref{fig:1}(c)]. After optical initialization, a square microwave pulse of variable duration $t$ is applied at frequency $\omega_d$ via the drive channel, with the central NV transition tuned to the cavity resonance. A subsequent weak cavity probe measures the spin-dependent cavity reflection, preferentially detecting spins near $\omega_c$. The signal exhibits pronounced oscillations as the pulse duration is varied: at a drive--cavity detuning $\Delta_{\mathrm{dc}}/(2\pi)=5\,\mathrm{MHz}$ and an intracavity driving strength $\Omega_{\mathrm{d}}/(2\pi)=0.53\,\mathrm{MHz}$, the oscillation frequency is approximately $5\,\mathrm{MHz}$ [Fig.~\ref{fig:1}(d)]. Here $\Omega_d$ is the Rabi drive amplitude set by the steady-state intracavity drive field. Intuitively, this single-pulse response would be attributed to a detuned Rabi oscillation at an effective Rabi frequency $\Omega_R=\sqrt{\Omega_d^2+\Delta_{\mathrm{dc}}^2}\approx|\Delta_{dc}|$ for $|\Delta_{\mathrm{dc}}|\gg\Omega_d$. However, the detuning-dependent Fourier spectra [Fig.~\ref{fig:1}(e)] show appreciable fringes persisting at large detuning, which is in contradiction to the detuned Rabi assumption that predicts a population-oscillation amplitude suppressed as $\Omega_d^2/(\Omega_d^2+\Delta_{\mathrm{dc}}^2)$, with the predicted spectrum shown in Fig.~\ref{fig:1}(g). The difference could be explained by the cavity transient fields, which convert the spin population to coherence and lead to an effective Ramsey interference as elaborated below.

\begin{figure*}[htbp]
    \centering
    \includegraphics[width=\textwidth]{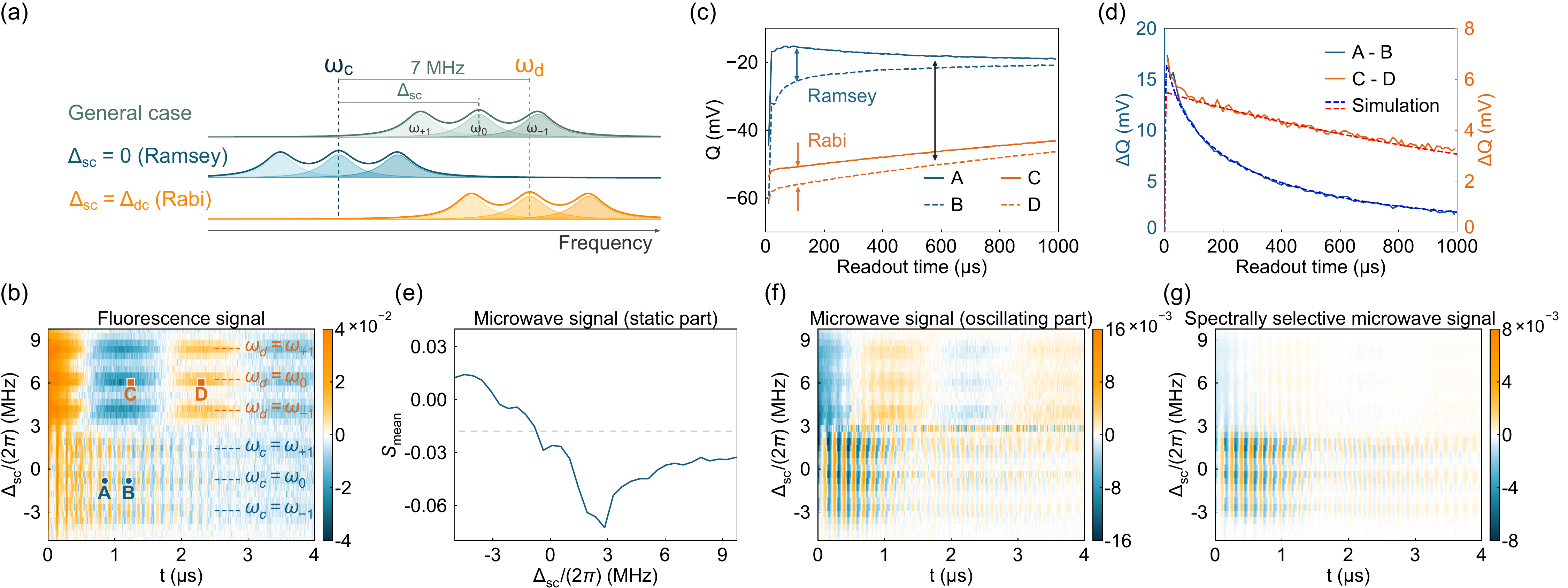}
    \caption{\textbf{Distinct roles of cavity transients and the continuous drive.} (a) Schematic spin profile of the NV ensemble relative to $\omega_c$ and $\omega_d$. 
    The spin-cavity detuning
    $\Delta_{\mathrm{sc}}=\omega_{s,0}-\omega_c$
    is defined with respect to the central ($m_I=0$) hyperfine transition. 
    (b) Measured fluorescence signals as functions of the evolution time $t$ and $\Delta_{\mathrm{sc}}$ with $\Delta_{\mathrm{dc}}/(2\pi)=7\,\mathrm{MHz}$ and $\Omega_d/(2\pi)=0.38\,\mathrm{MHz}$. The horizontal dashed lines mark the resonance frequencies of the three hyperfine transitions. 
    (c) Microwave quadrature ($Q$) traces for the four points A--D marked in (b).
    (d) Differences between the corresponding readout traces in (c). The dashed lines show simulations with $T_1=1.8\,\mathrm{ms}$, $T_\phi=3\,\upmu\mathrm{s}$, and a steady-state Rabi frequency of $\Omega_p/(2\pi)=0.032\,\mathrm{MHz}$. Each spin evolves under the probe field according to its detuning, and the total signal is calculated by averaging over the ensemble. The simulated curves are rescaled in amplitude to match the experimental data.
    (e--g) Measured microwave reflection signals. (e,f) Background and oscillating components obtained by averaging each readout trace from 20 to 560 $\upmu$s and separating the row mean. (g) Processed microwave signal obtained by subtracting the endpoint value from the trace average, emphasizing spins near the cavity resonance.
    All heatmaps are shown after subtraction of the row mean.
    }
    \label{fig:2}
\end{figure*}

The Ramsey character of these oscillations follows from the two spin rotations supplied by the cavity response at switch-on and switch-off. Right after an abrupt change of the external drive, the cavity field generates a decaying transient before reaching the new steady state. Although the cavity field in its steady state has only a single frequency component given by the external drive $\omega_d$, during the transient process a stimulated field component appears at the cavity frequency $\omega_c$. For a square pulse of an external drive $\eta e^{-i\omega_d t_1}$ $(0<t_1<t)$ with a duration $t$ applied to an initially empty cavity, the complex intracavity field amplitude neglecting the spin backaction in the laboratory frame is,
\begin{align}
\alpha_{\rm on}(t_1)&=\bar\alpha\left(e^{-i\omega_d t_1}-e^{-\kappa t_1/2}e^{-i\omega_c t_1}\right),
\label{eq:main_cavity_on}\\
\alpha_{\rm off}(t_2+t)&=\alpha_{\rm on}(t)e^{-\kappa t_2/2}e^{-i\omega_c t_2},
\label{eq:main_cavity_off}
\end{align}
where $0\leq t_1\leq t$ is the elapsed time from switch-on, $t_2\geq0$ is the elapsed time from switch-off, and $\bar\alpha=\eta/(\kappa/2-i\Delta_{\mathrm{dc}})$ is the steady-state cavity field. At switch-off, the cavity starts from the already stored field $\alpha_{\rm on}(t)$ and rings down with an oscillation frequency $\omega_c$. Both switching edges therefore generate cavity-frequency transients capable of rotating cavity-resonant spins even when the applied drive is far detuned from them.

For sufficiently large drive detuning, the $\omega_d$ component in Eq.~\eqref{eq:main_cavity_on} has little direct effect on spins near the cavity resonance, whose rotations are instead dominated by the $\omega_c$ transients. When the pulse duration is longer than the transient decay timescale such that $e^{-\kappa t/2}\ll1$, $\alpha_{\mathrm{on}}(t)\approx\Bar{\alpha}e^{-i\omega_d t}$, the two switching edges supply matching envelopes and thus coherent rotations of equal magnitude, with a relative phase set by $\Phi(t)=(\omega_d-\omega_s)t$ with $\omega_s$ the NV frequency. Cavity readout preferentially selects spins with $\omega_s\simeq\omega_c$, giving Ramsey fringes at $|\omega_d-\omega_c|$ [Fig.~\ref{fig:1}(d)]. Importantly, the transients need not be ideal $\pi/2$ pulses or exactly resonant with the spins: the first need only create a coherent superposition, and the second convert its relative phase into a population signal. The broadened ensemble thus allows appreciable interference from spins near the cavity resonance over a range of drive detunings as shown in Fig.~\ref{fig:1}(e), with the claim supported numerically in Fig.~\ref{fig:1}(f). 

We note that the Ramsey fringes in Fig.~\ref{fig:1}(d) decay with a fitted time constant of $0.65\,\upmu\mathrm{s}$, about three times the bare ensemble dephasing time $T_2^*\simeq0.21\,\upmu\mathrm{s}$. Such a prolonged coherence is caused by the finite duration of the first rotation with a cavity field ring-down timescale of $2/\kappa\sim 1.27\,\upmu\mathrm{s}$. 
During the nominal phase-accumulation interval, the decaying cavity field continuously converts a residual population imbalance into transverse spin coherence, effectively injecting coherence from the cavity field to spins. Theoretical analysis shows that this driven coherence time is set by $2/\kappa$, the same as the cavity field ring-down timescale (see detailed analysis in the supplement~\cite{SOM}).

\textit{Microwave readout with spectral resolution.---} To identify which spins produce the Ramsey fringes, we tune the NV hyperfine triplet across the cavity and drive frequencies while keeping their separation fixed at $\Delta_{\mathrm{dc}}/(2\pi)=7\,\mathrm{MHz}$ [Fig.~\ref{fig:2}(a)]. We track the triplet by the detuning $\Delta_{\mathrm{sc}}=\omega_{s,0}-\omega_c$ of its central transition. Fluorescence detects population changes across the ensemble and reveals two distinct responses as the control-pulse duration is varied [Fig.~\ref{fig:2}(b)]. Rapid Ramsey fringes with an oscillation frequency $\Delta_{\mathrm{dc}}$ appear when a hyperfine transition crosses the cavity resonance, where the switching transients rotate the spins. Slower Rabi oscillations with an oscillation frequency $\Omega_d$ appear when a transition crosses the drive frequency, where the applied field directly drives the spins resonantly.

Fluorescence readout gives approximately equal weight to population changes across the optically addressed spin spectrum. With microwave readout, the spin-cavity detuning affects both how sensitive the cavity responds to the spin population and how long the response persists upon the application of the cavity probe (population decay under the probe). The microwave readout is implemented by applying a weak probe at the cavity frequency and recording its reflected quadrature $Q$ as a function of readout time [Fig.~\ref{fig:2}(c)]. To isolate the population contrast due to Ramsey or Rabi oscillations, we compare two control-pulse durations corresponding to a signal peak and a valley at the same detuning: A and B in the Ramsey region, and C and D in the Rabi region [Fig.~\ref{fig:2}(b)]. The Ramsey contrast $|Q_A-Q_B|$ falls substantially during the first few hundred microseconds of readout, whereas the Rabi contrast $|Q_C-Q_D|$ remains appreciable at $1\,\mathrm{ms}$ [Fig.~\ref{fig:2}(d)] due to the larger spin-cavity detuning. Simulations of the coupled cavity-spin dynamics under the probe reproduce this difference in the time dependence of the two signals (dashed lines).

Such detuning-dependent cavity response is then used to choose which part of the spin spectrum to emphasize in the microwave readout. We select a readout time window from 20 to 560 $\upmu\mathrm{s}$ such that significant population decay happens for spins near cavity resonance while the population remains relatively unchanged for far-detuned spins. For Figs.~\ref{fig:2}(e,f), we plot the evolution signal under each $ \Delta_{sc}$ using the mean value of $Q$ in the selected trace window as each data point. The evolution signal is separated into the background part (e) that is independent of control-pulse duration, and the oscillating part (f) that evolves with time. Spins far from both $\omega_c$ and $\omega_d$ remain largely undriven and contribute to the background through their effect on the reflected probe. As the spin spectrum is tuned across the cavity, this contribution varies with detuning and accounts in part for the offset between the two pairs of traces marked by the black arrow in Fig.~\ref{fig:2}(c). For each point in Fig.~\ref{fig:2}(g), the data is obtained by calculating the mean value of $Q(t)-Q(560~\upmu \mathrm{s})$, which emphasizes the cavity-resonant Ramsey fringes and suppresses the contribution from far-detuned spins where the Rabi oscillation resides. A detailed numerical simulation is included in the supplement, which reproduces these experimental results~\cite{SOM}.

\begin{figure}[t]
    \centering
    \includegraphics[width=0.495\textwidth]{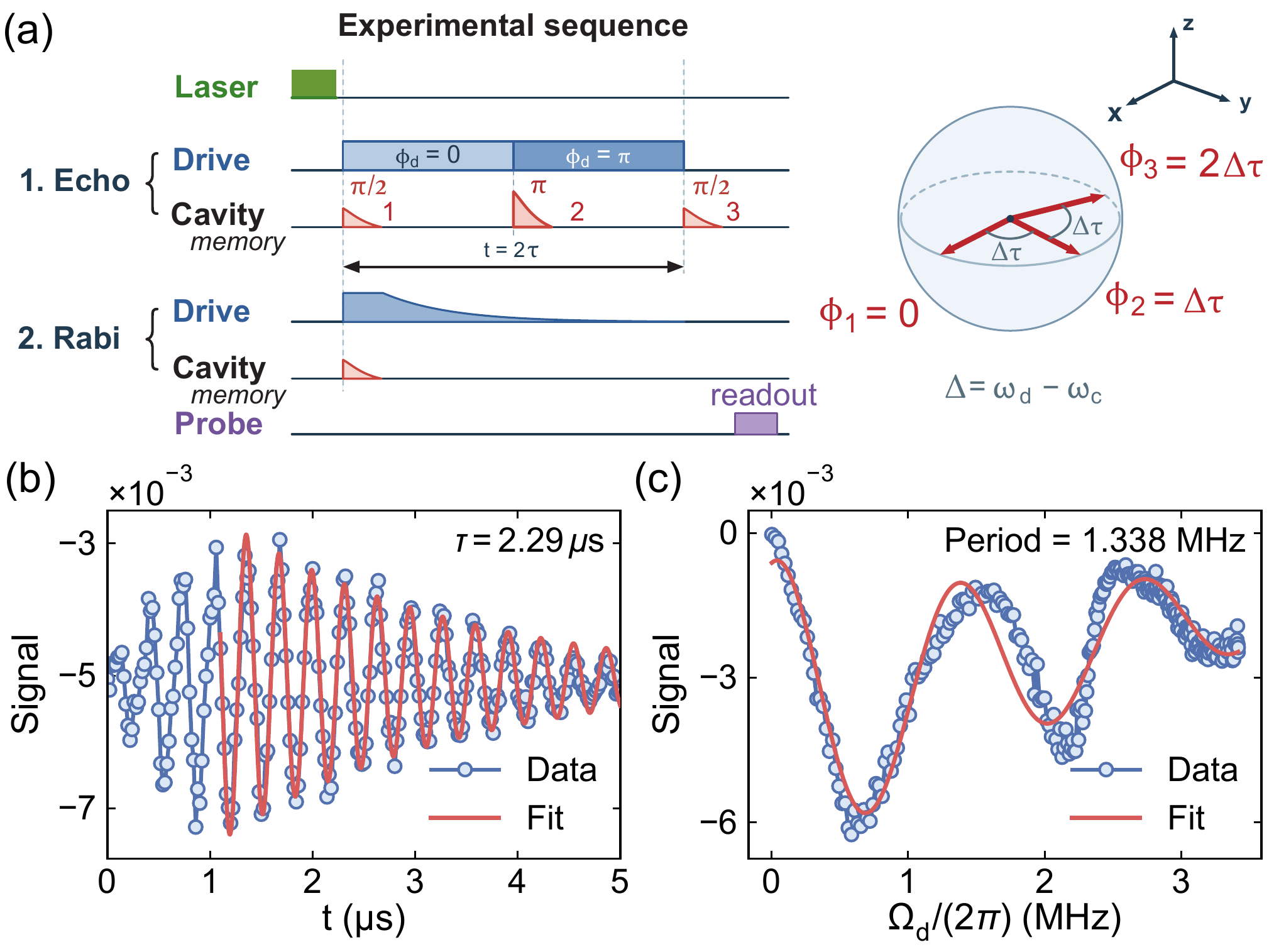}
    \caption{\textbf{Spin echo and Rabi control.} (a) Experimental control sequences (left) and the corresponding spin-echo rotation axes (right). 
    (b) Measured microwave reflection signal as a function of the total echo evolution time $t$, with $\Delta_{\mathrm{dc}}/(2\pi)=2.2\,\mathrm{MHz}$. The red solid line is a fit to an exponentially decaying oscillation model with a decay time constant $\tau=2.29\,\upmu\mathrm{s}$. (c) Population oscillations as a function of the drive strength $\Omega_d/(2\pi)$. The experiment is performed at $\Delta_{\mathrm{dc}}/(2\pi)=1\,\mathrm{MHz}$, using a $2\,\upmu\mathrm{s}$ driving pulse followed by an exponential tail with time constant $t_c=3\,\upmu\mathrm{s}$, truncated at $5t_c$ as shown in (a). The red curve is a phenomenological fit. }
    \label{fig:3}
\end{figure}

\textit{Quantum control by modulating the drive envelope.---} The transient-generated rotations can also be modified to construct other coherent operations. The same mechanism applies to any abrupt change of the drive field. Writing the external drive as $\eta q(t_1)e^{-i\omega_d t_1}$, a jump $\delta q_j=q(t_j^+)-q(t_j^-)$ at time $t_j$ adds a decaying cavity-field component 
$\delta\alpha_j(t_j+t_2)=-\bar\alpha\,\delta q_j e^{-i\omega_d t_j}
e^{-\kappa t_2/2}e^{-i\omega_c t_2}$,
where $t_2\geq0$ is the time since the jump. The magnitude and the complex phase of the discontinuity $\delta q_j e^{-i\omega_d t_j}$ sets the strength and transverse axis of the resulting spin rotation.
We first realize a spin-echo by inserting a $\pi$ phase jump halfway through the applied pulse [Fig.~\ref{fig:3}(a)]. The three discontinuities $\delta q_j$ given by the switch-on, phase-jump, switch-off have magnitudes in the ratio $1:2:1$. For well-separated transients, the middle event therefore supplies twice the rotation angle of either edge event, giving a nominal $\pi/2$-$\pi$-$\pi/2$ sequence when the drive amplitude is appropriately chosen. The transverse axis of the $\pi$ pulse inherits the drive-cavity phase accumulated at the switching time, as illustrated in Fig.~\ref{fig:3}(a). At $\Delta_{\mathrm{dc}}/(2\pi)=2.2\,\mathrm{MHz}$, the measured microwave response exhibits oscillations at a frequency $\Delta_{\mathrm{dc}}/2$ with a decay time $\tau=2.29\,\upmu\mathrm{s}$  [Fig.~\ref{fig:3}(b)].  The observed oscillation is caused by the time-dependence of the $\pi$ pulse rotation axis together with imperfect rotation angles. The latter effect is unavoidable even if a perfect $\pi/2$ condition is well calibrated for cavity-resonant spins due to the inhomogeneous feature of the spin ensemble, as discussed in detail in the supplement~\cite{SOM}.

Conversely, a single transient-induced rotation can be approximately isolated by retaining the abrupt switch-on and replacing the abrupt switch-off with a smooth ramp [Fig.~\ref{fig:3}(a)]. The gradual removal of the detuned drive reduces the final cavity-frequency transient and hence the second rotation. In Fig.~\ref{fig:3}(c), we use a $2\,\upmu\mathrm{s}$ pulse followed by an exponential tail with time constant $t_c=3\,\upmu\mathrm{s}$ at $\Delta_{\mathrm{dc}}/(2\pi)=1\,\mathrm{MHz}$. The measured population-dependent signal oscillates as the drive amplitude is varied, demonstrating coherent control through the strength of the initial transient. At larger amplitudes, $\Omega_d$ becomes comparable to or exceeds the drive-spin detuning, so the contribution of the drive field itself is not negligible anymore and the observed population signal need not follow the constant-period prediction for an isolated cavity transient.

\textit{Discussion and conclusion.---} The present experiment operates in a weak spin-backaction regime. Optical polarization covers approximately $1/9$ of the diamond volume, and the measured cavity linewidth broadening places an upper bound of about $0.15$ on the collective cooperativity. We can therefore describe the spin rotations using the cavity response to the applied pulse alone. Polarizing a larger fraction of the NV centers in the diamond would increase the number of participating spins and could bring the cooperativity close to unity. In that regime, the evolving spin polarization could appreciably modify the intracavity field that drives subsequent rotations. Further exploration of complicated dynamics of strongly-coupled NV-cQED systems with backaction as well as their applications in quantum control and sensing is of interest to future studies.

Even in this weak-backaction regime, the finite duration and phase of the transients shape the interference. As the first transient decays, it continues to generate transverse spin coherence during the nominal free-evolution interval, allowing the Ramsey fringes to persist beyond the bare ensemble dephasing time. The phase of each transient provides an additional control parameter: e.g., shifting one switching event relative to the others by a quarter of the drive-cavity beat period rotates its transverse control axis by $90^\circ$ for cavity-resonant spins. Such timing shifts could extend the demonstrated echo to more complex dynamical decoupling and sensing sequences.  Further exploring the cavity protection effects in hybrid quantum systems combining with cavity-generated dynamical decoupling and investigating their interplay with polariton hybridization and many-body dynamics are interesting future directions~\cite{lenz2021room,lei_manybody_2023}.  %

Instead of treating cavity transients as harmful effects, our work establishes cavity memory as a useful resource for coherent spin control. By shaping the envelope of a detuned microwave drive, we use the cavity response to generate coherent rotations needed for Ramsey interferometry, spin echo, and Rabi control. Furthermore, we demonstrate spectrally selective microwave readout at a single probe frequency by distinguishing the faster readout response of spins near the cavity resonance from the slower response of spins far-detuned from the cavity. This selectivity raises the prospect of using distinct spectral groups as parallel sensing channels and combining their signals through a single microwave probe. Combined with suitable cavity-mediated interactions such as one-axis twisting, coherent rotations could also support the generation of spin-squeezed states for entanglement-enhanced sensing~\cite{leroux_implementation_2010,fukumori_cavitymediated_2026a}. 

\acknowledgements
We acknowledge Gangqin Liu for fruitful discussions. This work was supported by the Fundamental Research Funds for the Central Universities, Peking University.

\bibliography{main} 

\begin{thebibliography}{36}%
\makeatletter
\providecommand \@ifxundefined [1]{%
 \@ifx{#1\undefined}
}%
\providecommand \@ifnum [1]{%
 \ifnum #1\expandafter \@firstoftwo
 \else \expandafter \@secondoftwo
 \fi
}%
\providecommand \@ifx [1]{%
 \ifx #1\expandafter \@firstoftwo
 \else \expandafter \@secondoftwo
 \fi
}%
\providecommand \natexlab [1]{#1}%
\providecommand \enquote  [1]{``#1''}%
\providecommand \bibnamefont  [1]{#1}%
\providecommand \bibfnamefont [1]{#1}%
\providecommand \citenamefont [1]{#1}%
\providecommand \href@noop [0]{\@secondoftwo}%
\providecommand \href [0]{\begingroup \@sanitize@url \@href}%
\providecommand \@href[1]{\@@startlink{#1}\@@href}%
\providecommand \@@href[1]{\endgroup#1\@@endlink}%
\providecommand \@sanitize@url [0]{\catcode `\\12\catcode `\$12\catcode `\&12\catcode `\#12\catcode `\^12\catcode `\_12\catcode `\%12\relax}%
\providecommand \@@startlink[1]{}%
\providecommand \@@endlink[0]{}%
\providecommand \url  [0]{\begingroup\@sanitize@url \@url }%
\providecommand \@url [1]{\endgroup\@href {#1}{\urlprefix }}%
\providecommand \urlprefix  [0]{URL }%
\providecommand \Eprint [0]{\href }%
\providecommand \doibase [0]{https://doi.org/}%
\providecommand \selectlanguage [0]{\@gobble}%
\providecommand \bibinfo  [0]{\@secondoftwo}%
\providecommand \bibfield  [0]{\@secondoftwo}%
\providecommand \translation [1]{[#1]}%
\providecommand \BibitemOpen [0]{}%
\providecommand \bibitemStop [0]{}%
\providecommand \bibitemNoStop [0]{.\EOS\space}%
\providecommand \EOS [0]{\spacefactor3000\relax}%
\providecommand \BibitemShut  [1]{\csname bibitem#1\endcsname}%
\let\auto@bib@innerbib\@empty
\bibitem [{\citenamefont {Kubo}\ \emph {et~al.}(2010)\citenamefont {Kubo}, \citenamefont {Ong}, \citenamefont {Bertet}, \citenamefont {Vion}, \citenamefont {Jacques}, \citenamefont {Zheng}, \citenamefont {Dr{\'e}au}, \citenamefont {Roch}, \citenamefont {Auffeves}, \citenamefont {Jelezko}, \citenamefont {Wrachtrup}, \citenamefont {Barthe}, \citenamefont {Bergonzo},\ and\ \citenamefont {Esteve}}]{kubo_strong_2010}%
  \BibitemOpen
  \bibfield  {author} {\bibinfo {author} {\bibfnamefont {Y.}~\bibnamefont {Kubo}}, \bibinfo {author} {\bibfnamefont {F.~R.}\ \bibnamefont {Ong}}, \bibinfo {author} {\bibfnamefont {P.}~\bibnamefont {Bertet}}, \bibinfo {author} {\bibfnamefont {D.}~\bibnamefont {Vion}}, \bibinfo {author} {\bibfnamefont {V.}~\bibnamefont {Jacques}}, \bibinfo {author} {\bibfnamefont {D.}~\bibnamefont {Zheng}}, \bibinfo {author} {\bibfnamefont {A.}~\bibnamefont {Dr{\'e}au}}, \bibinfo {author} {\bibfnamefont {J.-F.}\ \bibnamefont {Roch}}, \bibinfo {author} {\bibfnamefont {A.}~\bibnamefont {Auffeves}}, \bibinfo {author} {\bibfnamefont {F.}~\bibnamefont {Jelezko}}, \bibinfo {author} {\bibfnamefont {J.}~\bibnamefont {Wrachtrup}}, \bibinfo {author} {\bibfnamefont {M.~F.}\ \bibnamefont {Barthe}}, \bibinfo {author} {\bibfnamefont {P.}~\bibnamefont {Bergonzo}},\ and\ \bibinfo {author} {\bibfnamefont {D.}~\bibnamefont {Esteve}},\ }\bibfield  {title} {\bibinfo {title} {Strong {{Coupling}} of a {{Spin Ensemble}} to a {{Superconducting
  Resonator}}},\ }\href {https://doi.org/10.1103/PhysRevLett.105.140502} {\bibfield  {journal} {\bibinfo  {journal} {Phys. Rev. Lett.}\ }\textbf {\bibinfo {volume} {105}},\ \bibinfo {pages} {140502} (\bibinfo {year} {2010})}\BibitemShut {NoStop}%
\bibitem [{\citenamefont {Schuster}\ \emph {et~al.}(2010)\citenamefont {Schuster}, \citenamefont {Sears}, \citenamefont {Ginossar}, \citenamefont {DiCarlo}, \citenamefont {Frunzio}, \citenamefont {Morton}, \citenamefont {Wu}, \citenamefont {Briggs}, \citenamefont {Buckley}, \citenamefont {Awschalom},\ and\ \citenamefont {Schoelkopf}}]{schuster_highcooperativity_2010}%
  \BibitemOpen
  \bibfield  {author} {\bibinfo {author} {\bibfnamefont {D.~I.}\ \bibnamefont {Schuster}}, \bibinfo {author} {\bibfnamefont {A.~P.}\ \bibnamefont {Sears}}, \bibinfo {author} {\bibfnamefont {E.}~\bibnamefont {Ginossar}}, \bibinfo {author} {\bibfnamefont {L.}~\bibnamefont {DiCarlo}}, \bibinfo {author} {\bibfnamefont {L.}~\bibnamefont {Frunzio}}, \bibinfo {author} {\bibfnamefont {J.~J.~L.}\ \bibnamefont {Morton}}, \bibinfo {author} {\bibfnamefont {H.}~\bibnamefont {Wu}}, \bibinfo {author} {\bibfnamefont {G.~A.~D.}\ \bibnamefont {Briggs}}, \bibinfo {author} {\bibfnamefont {B.~B.}\ \bibnamefont {Buckley}}, \bibinfo {author} {\bibfnamefont {D.~D.}\ \bibnamefont {Awschalom}},\ and\ \bibinfo {author} {\bibfnamefont {R.~J.}\ \bibnamefont {Schoelkopf}},\ }\bibfield  {title} {\bibinfo {title} {High-{{Cooperativity Coupling}} of {{Electron-Spin Ensembles}} to {{Superconducting Cavities}}},\ }\href {https://doi.org/10.1103/PhysRevLett.105.140501} {\bibfield  {journal} {\bibinfo  {journal} {Phys. Rev. Lett.}\ }\textbf
  {\bibinfo {volume} {105}},\ \bibinfo {pages} {140501} (\bibinfo {year} {2010})}\BibitemShut {NoStop}%
\bibitem [{\citenamefont {Ams{\"u}ss}\ \emph {et~al.}(2011)\citenamefont {Ams{\"u}ss}, \citenamefont {Koller}, \citenamefont {N{\"o}bauer}, \citenamefont {Putz}, \citenamefont {Rotter}, \citenamefont {Sandner}, \citenamefont {Schneider}, \citenamefont {Schramb{\"o}ck}, \citenamefont {Steinhauser}, \citenamefont {Ritsch}, \citenamefont {Schmiedmayer},\ and\ \citenamefont {Majer}}]{amsuss_cavity_2011}%
  \BibitemOpen
  \bibfield  {author} {\bibinfo {author} {\bibfnamefont {R.}~\bibnamefont {Ams{\"u}ss}}, \bibinfo {author} {\bibfnamefont {{\relax Ch}.}~\bibnamefont {Koller}}, \bibinfo {author} {\bibfnamefont {T.}~\bibnamefont {N{\"o}bauer}}, \bibinfo {author} {\bibfnamefont {S.}~\bibnamefont {Putz}}, \bibinfo {author} {\bibfnamefont {S.}~\bibnamefont {Rotter}}, \bibinfo {author} {\bibfnamefont {K.}~\bibnamefont {Sandner}}, \bibinfo {author} {\bibfnamefont {S.}~\bibnamefont {Schneider}}, \bibinfo {author} {\bibfnamefont {M.}~\bibnamefont {Schramb{\"o}ck}}, \bibinfo {author} {\bibfnamefont {G.}~\bibnamefont {Steinhauser}}, \bibinfo {author} {\bibfnamefont {H.}~\bibnamefont {Ritsch}}, \bibinfo {author} {\bibfnamefont {J.}~\bibnamefont {Schmiedmayer}},\ and\ \bibinfo {author} {\bibfnamefont {J.}~\bibnamefont {Majer}},\ }\bibfield  {title} {\bibinfo {title} {Cavity {{QED}} with {{Magnetically Coupled Collective Spin States}}},\ }\href {https://doi.org/10.1103/PhysRevLett.107.060502} {\bibfield  {journal} {\bibinfo  {journal}
  {Phys. Rev. Lett.}\ }\textbf {\bibinfo {volume} {107}},\ \bibinfo {pages} {060502} (\bibinfo {year} {2011})}\BibitemShut {NoStop}%
\bibitem [{\citenamefont {Probst}\ \emph {et~al.}(2013)\citenamefont {Probst}, \citenamefont {Rotzinger}, \citenamefont {W{\"u}nsch}, \citenamefont {Jung}, \citenamefont {Jerger}, \citenamefont {Siegel}, \citenamefont {Ustinov},\ and\ \citenamefont {Bushev}}]{probst_anisotropic_2013}%
  \BibitemOpen
  \bibfield  {author} {\bibinfo {author} {\bibfnamefont {S.}~\bibnamefont {Probst}}, \bibinfo {author} {\bibfnamefont {H.}~\bibnamefont {Rotzinger}}, \bibinfo {author} {\bibfnamefont {S.}~\bibnamefont {W{\"u}nsch}}, \bibinfo {author} {\bibfnamefont {P.}~\bibnamefont {Jung}}, \bibinfo {author} {\bibfnamefont {M.}~\bibnamefont {Jerger}}, \bibinfo {author} {\bibfnamefont {M.}~\bibnamefont {Siegel}}, \bibinfo {author} {\bibfnamefont {A.~V.}\ \bibnamefont {Ustinov}},\ and\ \bibinfo {author} {\bibfnamefont {P.~A.}\ \bibnamefont {Bushev}},\ }\bibfield  {title} {\bibinfo {title} {Anisotropic {{Rare-Earth Spin Ensemble Strongly Coupled}} to a {{Superconducting Resonator}}},\ }\href {https://doi.org/10.1103/PhysRevLett.110.157001} {\bibfield  {journal} {\bibinfo  {journal} {Phys. Rev. Lett.}\ }\textbf {\bibinfo {volume} {110}},\ \bibinfo {pages} {157001} (\bibinfo {year} {2013})}\BibitemShut {NoStop}%
\bibitem [{\citenamefont {Ranjan}\ \emph {et~al.}(2013)\citenamefont {Ranjan}, \citenamefont {{de Lange}}, \citenamefont {Schutjens}, \citenamefont {Debelhoir}, \citenamefont {Groen}, \citenamefont {Szombati}, \citenamefont {Thoen}, \citenamefont {Klapwijk}, \citenamefont {Hanson},\ and\ \citenamefont {DiCarlo}}]{ranjan_probing_2013}%
  \BibitemOpen
  \bibfield  {author} {\bibinfo {author} {\bibfnamefont {V.}~\bibnamefont {Ranjan}}, \bibinfo {author} {\bibfnamefont {G.}~\bibnamefont {{de Lange}}}, \bibinfo {author} {\bibfnamefont {R.}~\bibnamefont {Schutjens}}, \bibinfo {author} {\bibfnamefont {T.}~\bibnamefont {Debelhoir}}, \bibinfo {author} {\bibfnamefont {J.~P.}\ \bibnamefont {Groen}}, \bibinfo {author} {\bibfnamefont {D.}~\bibnamefont {Szombati}}, \bibinfo {author} {\bibfnamefont {D.~J.}\ \bibnamefont {Thoen}}, \bibinfo {author} {\bibfnamefont {T.~M.}\ \bibnamefont {Klapwijk}}, \bibinfo {author} {\bibfnamefont {R.}~\bibnamefont {Hanson}},\ and\ \bibinfo {author} {\bibfnamefont {L.}~\bibnamefont {DiCarlo}},\ }\bibfield  {title} {\bibinfo {title} {Probing {{Dynamics}} of an {{Electron-Spin Ensemble}} via a {{Superconducting Resonator}}},\ }\href {https://doi.org/10.1103/PhysRevLett.110.067004} {\bibfield  {journal} {\bibinfo  {journal} {Phys. Rev. Lett.}\ }\textbf {\bibinfo {volume} {110}},\ \bibinfo {pages} {067004} (\bibinfo {year}
  {2013})}\BibitemShut {NoStop}%
\bibitem [{\citenamefont {Zhu}\ \emph {et~al.}(2011)\citenamefont {Zhu}, \citenamefont {Saito}, \citenamefont {Kemp}, \citenamefont {Kakuyanagi}, \citenamefont {Karimoto}, \citenamefont {Nakano}, \citenamefont {Munro}, \citenamefont {Tokura}, \citenamefont {Everitt}, \citenamefont {Nemoto}, \citenamefont {Kasu}, \citenamefont {Mizuochi},\ and\ \citenamefont {Semba}}]{zhu_coherent_2011}%
  \BibitemOpen
  \bibfield  {author} {\bibinfo {author} {\bibfnamefont {X.}~\bibnamefont {Zhu}}, \bibinfo {author} {\bibfnamefont {S.}~\bibnamefont {Saito}}, \bibinfo {author} {\bibfnamefont {A.}~\bibnamefont {Kemp}}, \bibinfo {author} {\bibfnamefont {K.}~\bibnamefont {Kakuyanagi}}, \bibinfo {author} {\bibfnamefont {S.-i.}\ \bibnamefont {Karimoto}}, \bibinfo {author} {\bibfnamefont {H.}~\bibnamefont {Nakano}}, \bibinfo {author} {\bibfnamefont {W.~J.}\ \bibnamefont {Munro}}, \bibinfo {author} {\bibfnamefont {Y.}~\bibnamefont {Tokura}}, \bibinfo {author} {\bibfnamefont {M.~S.}\ \bibnamefont {Everitt}}, \bibinfo {author} {\bibfnamefont {K.}~\bibnamefont {Nemoto}}, \bibinfo {author} {\bibfnamefont {M.}~\bibnamefont {Kasu}}, \bibinfo {author} {\bibfnamefont {N.}~\bibnamefont {Mizuochi}},\ and\ \bibinfo {author} {\bibfnamefont {K.}~\bibnamefont {Semba}},\ }\bibfield  {title} {\bibinfo {title} {Coherent coupling of a superconducting flux qubit to an electron spin ensemble in diamond},\ }\href {https://doi.org/10.1038/nature10462}
  {\bibfield  {journal} {\bibinfo  {journal} {Nature}\ }\textbf {\bibinfo {volume} {478}},\ \bibinfo {pages} {221} (\bibinfo {year} {2011})}\BibitemShut {NoStop}%
\bibitem [{\citenamefont {Grezes}\ \emph {et~al.}(2014)\citenamefont {Grezes}, \citenamefont {Julsgaard}, \citenamefont {Kubo}, \citenamefont {Stern}, \citenamefont {Umeda}, \citenamefont {Isoya}, \citenamefont {Sumiya}, \citenamefont {Abe}, \citenamefont {Onoda}, \citenamefont {Ohshima}, \citenamefont {Jacques}, \citenamefont {Esteve}, \citenamefont {Vion}, \citenamefont {Esteve}, \citenamefont {M{\o}lmer},\ and\ \citenamefont {Bertet}}]{grezes_multimode_2014}%
  \BibitemOpen
  \bibfield  {author} {\bibinfo {author} {\bibfnamefont {C.}~\bibnamefont {Grezes}}, \bibinfo {author} {\bibfnamefont {B.}~\bibnamefont {Julsgaard}}, \bibinfo {author} {\bibfnamefont {Y.}~\bibnamefont {Kubo}}, \bibinfo {author} {\bibfnamefont {M.}~\bibnamefont {Stern}}, \bibinfo {author} {\bibfnamefont {T.}~\bibnamefont {Umeda}}, \bibinfo {author} {\bibfnamefont {J.}~\bibnamefont {Isoya}}, \bibinfo {author} {\bibfnamefont {H.}~\bibnamefont {Sumiya}}, \bibinfo {author} {\bibfnamefont {H.}~\bibnamefont {Abe}}, \bibinfo {author} {\bibfnamefont {S.}~\bibnamefont {Onoda}}, \bibinfo {author} {\bibfnamefont {T.}~\bibnamefont {Ohshima}}, \bibinfo {author} {\bibfnamefont {V.}~\bibnamefont {Jacques}}, \bibinfo {author} {\bibfnamefont {J.}~\bibnamefont {Esteve}}, \bibinfo {author} {\bibfnamefont {D.}~\bibnamefont {Vion}}, \bibinfo {author} {\bibfnamefont {D.}~\bibnamefont {Esteve}}, \bibinfo {author} {\bibfnamefont {K.}~\bibnamefont {M{\o}lmer}},\ and\ \bibinfo {author} {\bibfnamefont {P.}~\bibnamefont {Bertet}},\
  }\bibfield  {title} {\bibinfo {title} {Multimode {{Storage}} and {{Retrieval}} of {{Microwave Fields}} in a {{Spin Ensemble}}},\ }\href {https://doi.org/10.1103/PhysRevX.4.021049} {\bibfield  {journal} {\bibinfo  {journal} {Phys. Rev. X}\ }\textbf {\bibinfo {volume} {4}},\ \bibinfo {pages} {021049} (\bibinfo {year} {2014})}\BibitemShut {NoStop}%
\bibitem [{\citenamefont {Putz}\ \emph {et~al.}(2014)\citenamefont {Putz}, \citenamefont {Krimer}, \citenamefont {Ams{\"u}ss}, \citenamefont {Valookaran}, \citenamefont {N{\"o}bauer}, \citenamefont {Schmiedmayer}, \citenamefont {Rotter},\ and\ \citenamefont {Majer}}]{putz_protecting_2014}%
  \BibitemOpen
  \bibfield  {author} {\bibinfo {author} {\bibfnamefont {S.}~\bibnamefont {Putz}}, \bibinfo {author} {\bibfnamefont {D.~O.}\ \bibnamefont {Krimer}}, \bibinfo {author} {\bibfnamefont {R.}~\bibnamefont {Ams{\"u}ss}}, \bibinfo {author} {\bibfnamefont {A.}~\bibnamefont {Valookaran}}, \bibinfo {author} {\bibfnamefont {T.}~\bibnamefont {N{\"o}bauer}}, \bibinfo {author} {\bibfnamefont {J.}~\bibnamefont {Schmiedmayer}}, \bibinfo {author} {\bibfnamefont {S.}~\bibnamefont {Rotter}},\ and\ \bibinfo {author} {\bibfnamefont {J.}~\bibnamefont {Majer}},\ }\bibfield  {title} {\bibinfo {title} {Protecting a spin ensemble against decoherence in the strong-coupling regime of cavity {{QED}}},\ }\href {https://doi.org/10.1038/nphys3050} {\bibfield  {journal} {\bibinfo  {journal} {Nat. Phys.}\ }\textbf {\bibinfo {volume} {10}},\ \bibinfo {pages} {720} (\bibinfo {year} {2014})}\BibitemShut {NoStop}%
\bibitem [{\citenamefont {Angerer}\ \emph {et~al.}(2018)\citenamefont {Angerer}, \citenamefont {Streltsov}, \citenamefont {Astner}, \citenamefont {Putz}, \citenamefont {Sumiya}, \citenamefont {Onoda}, \citenamefont {Isoya}, \citenamefont {Munro}, \citenamefont {Nemoto}, \citenamefont {Schmiedmayer},\ and\ \citenamefont {Majer}}]{angerer_superradiant_2018}%
  \BibitemOpen
  \bibfield  {author} {\bibinfo {author} {\bibfnamefont {A.}~\bibnamefont {Angerer}}, \bibinfo {author} {\bibfnamefont {K.}~\bibnamefont {Streltsov}}, \bibinfo {author} {\bibfnamefont {T.}~\bibnamefont {Astner}}, \bibinfo {author} {\bibfnamefont {S.}~\bibnamefont {Putz}}, \bibinfo {author} {\bibfnamefont {H.}~\bibnamefont {Sumiya}}, \bibinfo {author} {\bibfnamefont {S.}~\bibnamefont {Onoda}}, \bibinfo {author} {\bibfnamefont {J.}~\bibnamefont {Isoya}}, \bibinfo {author} {\bibfnamefont {W.~J.}\ \bibnamefont {Munro}}, \bibinfo {author} {\bibfnamefont {K.}~\bibnamefont {Nemoto}}, \bibinfo {author} {\bibfnamefont {J.}~\bibnamefont {Schmiedmayer}},\ and\ \bibinfo {author} {\bibfnamefont {J.}~\bibnamefont {Majer}},\ }\bibfield  {title} {\bibinfo {title} {Superradiant emission from colour centres in diamond},\ }\href {https://doi.org/10.1038/s41567-018-0269-7} {\bibfield  {journal} {\bibinfo  {journal} {Nat. Phys.}\ }\textbf {\bibinfo {volume} {14}},\ \bibinfo {pages} {1168} (\bibinfo {year} {2018})}\BibitemShut
  {NoStop}%
\bibitem [{\citenamefont {Weichselbaumer}\ \emph {et~al.}(2020)\citenamefont {Weichselbaumer}, \citenamefont {Zens}, \citenamefont {Zollitsch}, \citenamefont {Brandt}, \citenamefont {Rotter}, \citenamefont {Gross},\ and\ \citenamefont {Huebl}}]{weichselbaumer_echo_2020}%
  \BibitemOpen
  \bibfield  {author} {\bibinfo {author} {\bibfnamefont {S.}~\bibnamefont {Weichselbaumer}}, \bibinfo {author} {\bibfnamefont {M.}~\bibnamefont {Zens}}, \bibinfo {author} {\bibfnamefont {C.~W.}\ \bibnamefont {Zollitsch}}, \bibinfo {author} {\bibfnamefont {M.~S.}\ \bibnamefont {Brandt}}, \bibinfo {author} {\bibfnamefont {S.}~\bibnamefont {Rotter}}, \bibinfo {author} {\bibfnamefont {R.}~\bibnamefont {Gross}},\ and\ \bibinfo {author} {\bibfnamefont {H.}~\bibnamefont {Huebl}},\ }\bibfield  {title} {\bibinfo {title} {Echo {{Trains}} in {{Pulsed Electron Spin Resonance}} of a {{Strongly Coupled Spin Ensemble}}},\ }\href {https://doi.org/10.1103/PhysRevLett.125.137701} {\bibfield  {journal} {\bibinfo  {journal} {Phys. Rev. Lett.}\ }\textbf {\bibinfo {volume} {125}},\ \bibinfo {pages} {137701} (\bibinfo {year} {2020})}\BibitemShut {NoStop}%
\bibitem [{\citenamefont {Debnath}\ \emph {et~al.}(2020)\citenamefont {Debnath}, \citenamefont {Dold}, \citenamefont {Morton},\ and\ \citenamefont {M{\o}lmer}}]{debnath_selfstimulated_2020}%
  \BibitemOpen
  \bibfield  {author} {\bibinfo {author} {\bibfnamefont {K.}~\bibnamefont {Debnath}}, \bibinfo {author} {\bibfnamefont {G.}~\bibnamefont {Dold}}, \bibinfo {author} {\bibfnamefont {J.~J.~L.}\ \bibnamefont {Morton}},\ and\ \bibinfo {author} {\bibfnamefont {K.}~\bibnamefont {M{\o}lmer}},\ }\bibfield  {title} {\bibinfo {title} {Self-{{Stimulated Pulse Echo Trains}} from {{Inhomogeneously Broadened Spin Ensembles}}},\ }\href {https://doi.org/10.1103/PhysRevLett.125.137702} {\bibfield  {journal} {\bibinfo  {journal} {Phys. Rev. Lett.}\ }\textbf {\bibinfo {volume} {125}},\ \bibinfo {pages} {137702} (\bibinfo {year} {2020})}\BibitemShut {NoStop}%
\bibitem [{\citenamefont {Kersten}\ \emph {et~al.}(2023)\citenamefont {Kersten}, \citenamefont {{de Zordo}}, \citenamefont {Diekmann}, \citenamefont {Reiter}, \citenamefont {Zens}, \citenamefont {Kanagin}, \citenamefont {Rotter}, \citenamefont {Schmiedmayer},\ and\ \citenamefont {Angerer}}]{kersten_triggered_2023}%
  \BibitemOpen
  \bibfield  {author} {\bibinfo {author} {\bibfnamefont {W.}~\bibnamefont {Kersten}}, \bibinfo {author} {\bibfnamefont {N.}~\bibnamefont {{de Zordo}}}, \bibinfo {author} {\bibfnamefont {O.}~\bibnamefont {Diekmann}}, \bibinfo {author} {\bibfnamefont {T.}~\bibnamefont {Reiter}}, \bibinfo {author} {\bibfnamefont {M.}~\bibnamefont {Zens}}, \bibinfo {author} {\bibfnamefont {A.~N.}\ \bibnamefont {Kanagin}}, \bibinfo {author} {\bibfnamefont {S.}~\bibnamefont {Rotter}}, \bibinfo {author} {\bibfnamefont {J.}~\bibnamefont {Schmiedmayer}},\ and\ \bibinfo {author} {\bibfnamefont {A.}~\bibnamefont {Angerer}},\ }\bibfield  {title} {\bibinfo {title} {Triggered {{Superradiance}} and {{Spin Inversion Storage}} in a {{Hybrid Quantum System}}},\ }\href {https://doi.org/10.1103/PhysRevLett.131.043601} {\bibfield  {journal} {\bibinfo  {journal} {Phys. Rev. Lett.}\ }\textbf {\bibinfo {volume} {131}},\ \bibinfo {pages} {043601} (\bibinfo {year} {2023})}\BibitemShut {NoStop}%
\bibitem [{\citenamefont {Kersten}\ \emph {et~al.}(2026)\citenamefont {Kersten}, \citenamefont {De~Zordo}, \citenamefont {Diekmann}, \citenamefont {Redchenko}, \citenamefont {Kanagin}, \citenamefont {Angerer}, \citenamefont {Munro}, \citenamefont {Nemoto}, \citenamefont {Mazets}, \citenamefont {Rotter}, \citenamefont {Pohl},\ and\ \citenamefont {Schmiedmayer}}]{kersten_selfinduced_2026}%
  \BibitemOpen
  \bibfield  {author} {\bibinfo {author} {\bibfnamefont {W.}~\bibnamefont {Kersten}}, \bibinfo {author} {\bibfnamefont {N.}~\bibnamefont {De~Zordo}}, \bibinfo {author} {\bibfnamefont {O.}~\bibnamefont {Diekmann}}, \bibinfo {author} {\bibfnamefont {E.~S.}\ \bibnamefont {Redchenko}}, \bibinfo {author} {\bibfnamefont {A.~N.}\ \bibnamefont {Kanagin}}, \bibinfo {author} {\bibfnamefont {A.}~\bibnamefont {Angerer}}, \bibinfo {author} {\bibfnamefont {W.~J.}\ \bibnamefont {Munro}}, \bibinfo {author} {\bibfnamefont {K.}~\bibnamefont {Nemoto}}, \bibinfo {author} {\bibfnamefont {I.~E.}\ \bibnamefont {Mazets}}, \bibinfo {author} {\bibfnamefont {S.}~\bibnamefont {Rotter}}, \bibinfo {author} {\bibfnamefont {T.}~\bibnamefont {Pohl}},\ and\ \bibinfo {author} {\bibfnamefont {J.}~\bibnamefont {Schmiedmayer}},\ }\bibfield  {title} {\bibinfo {title} {Self-induced superradiant masing},\ }\href {https://doi.org/10.1038/s41567-025-03123-0} {\bibfield  {journal} {\bibinfo  {journal} {Nat. Phys.}\ }\textbf {\bibinfo {volume} {22}},\
  \bibinfo {pages} {158} (\bibinfo {year} {2026})}\BibitemShut {NoStop}%
\bibitem [{\citenamefont {Wu}\ \emph {et~al.}(2026)\citenamefont {Wu}, \citenamefont {Xiao}, \citenamefont {Wang}, \citenamefont {Chen}, \citenamefont {Wang}, \citenamefont {Ai}, \citenamefont {Li},\ and\ \citenamefont {Liu}}]{wu_sustained_2026}%
  \BibitemOpen
  \bibfield  {author} {\bibinfo {author} {\bibfnamefont {W.-J.}\ \bibnamefont {Wu}}, \bibinfo {author} {\bibfnamefont {D.-W.}\ \bibnamefont {Xiao}}, \bibinfo {author} {\bibfnamefont {W.-T.}\ \bibnamefont {Wang}}, \bibinfo {author} {\bibfnamefont {X.-Y.}\ \bibnamefont {Chen}}, \bibinfo {author} {\bibfnamefont {X.-F.}\ \bibnamefont {Wang}}, \bibinfo {author} {\bibfnamefont {M.-Z.}\ \bibnamefont {Ai}}, \bibinfo {author} {\bibfnamefont {Q.}~\bibnamefont {Li}},\ and\ \bibinfo {author} {\bibfnamefont {R.-B.}\ \bibnamefont {Liu}},\ }\href@noop {} {\bibinfo {title} {Sustained macroscopic quantum coherence in a superradiant solid under ambient conditions}} (\bibinfo {year} {2026}),\ \Eprint {https://arxiv.org/abs/2608.23748} {arXiv:2608.23748 [quant-ph]} \BibitemShut {NoStop}%
\bibitem [{\citenamefont {Lei}\ \emph {et~al.}(2023)\citenamefont {Lei}, \citenamefont {Fukumori}, \citenamefont {Rochman}, \citenamefont {Zhu}, \citenamefont {Endres}, \citenamefont {Choi},\ and\ \citenamefont {Faraon}}]{lei_manybody_2023}%
  \BibitemOpen
  \bibfield  {author} {\bibinfo {author} {\bibfnamefont {M.}~\bibnamefont {Lei}}, \bibinfo {author} {\bibfnamefont {R.}~\bibnamefont {Fukumori}}, \bibinfo {author} {\bibfnamefont {J.}~\bibnamefont {Rochman}}, \bibinfo {author} {\bibfnamefont {B.}~\bibnamefont {Zhu}}, \bibinfo {author} {\bibfnamefont {M.}~\bibnamefont {Endres}}, \bibinfo {author} {\bibfnamefont {J.}~\bibnamefont {Choi}},\ and\ \bibinfo {author} {\bibfnamefont {A.}~\bibnamefont {Faraon}},\ }\bibfield  {title} {\bibinfo {title} {Many-body cavity quantum electrodynamics with driven inhomogeneous emitters},\ }\href {https://doi.org/10.1038/s41586-023-05884-1} {\bibfield  {journal} {\bibinfo  {journal} {Nature}\ }\textbf {\bibinfo {volume} {617}},\ \bibinfo {pages} {271} (\bibinfo {year} {2023})}\BibitemShut {NoStop}%
\bibitem [{\citenamefont {Bienfait}\ \emph {et~al.}(2016)\citenamefont {Bienfait}, \citenamefont {Pla}, \citenamefont {Kubo}, \citenamefont {Stern}, \citenamefont {Zhou}, \citenamefont {Lo}, \citenamefont {Weis}, \citenamefont {Schenkel}, \citenamefont {Thewalt}, \citenamefont {Vion}, \citenamefont {Esteve}, \citenamefont {Julsgaard}, \citenamefont {M{\o}lmer}, \citenamefont {Morton},\ and\ \citenamefont {Bertet}}]{bienfait_reaching_2016}%
  \BibitemOpen
  \bibfield  {author} {\bibinfo {author} {\bibfnamefont {A.}~\bibnamefont {Bienfait}}, \bibinfo {author} {\bibfnamefont {J.~J.}\ \bibnamefont {Pla}}, \bibinfo {author} {\bibfnamefont {Y.}~\bibnamefont {Kubo}}, \bibinfo {author} {\bibfnamefont {M.}~\bibnamefont {Stern}}, \bibinfo {author} {\bibfnamefont {X.}~\bibnamefont {Zhou}}, \bibinfo {author} {\bibfnamefont {C.~C.}\ \bibnamefont {Lo}}, \bibinfo {author} {\bibfnamefont {C.~D.}\ \bibnamefont {Weis}}, \bibinfo {author} {\bibfnamefont {T.}~\bibnamefont {Schenkel}}, \bibinfo {author} {\bibfnamefont {M.~L.~W.}\ \bibnamefont {Thewalt}}, \bibinfo {author} {\bibfnamefont {D.}~\bibnamefont {Vion}}, \bibinfo {author} {\bibfnamefont {D.}~\bibnamefont {Esteve}}, \bibinfo {author} {\bibfnamefont {B.}~\bibnamefont {Julsgaard}}, \bibinfo {author} {\bibfnamefont {K.}~\bibnamefont {M{\o}lmer}}, \bibinfo {author} {\bibfnamefont {J.~J.~L.}\ \bibnamefont {Morton}},\ and\ \bibinfo {author} {\bibfnamefont {P.}~\bibnamefont {Bertet}},\ }\bibfield  {title} {\bibinfo {title}
  {Reaching the quantum limit of sensitivity in electron spin resonance},\ }\href {https://doi.org/10.1038/nnano.2015.282} {\bibfield  {journal} {\bibinfo  {journal} {Nat. Nanotechnol.}\ }\textbf {\bibinfo {volume} {11}},\ \bibinfo {pages} {253} (\bibinfo {year} {2016})}\BibitemShut {NoStop}%
\bibitem [{\citenamefont {Ebel}\ \emph {et~al.}(2021)\citenamefont {Ebel}, \citenamefont {Joas}, \citenamefont {Schalk}, \citenamefont {Weinbrenner}, \citenamefont {Angerer}, \citenamefont {Majer},\ and\ \citenamefont {Reinhard}}]{ebel_dispersive_2021}%
  \BibitemOpen
  \bibfield  {author} {\bibinfo {author} {\bibfnamefont {J.}~\bibnamefont {Ebel}}, \bibinfo {author} {\bibfnamefont {T.}~\bibnamefont {Joas}}, \bibinfo {author} {\bibfnamefont {M.}~\bibnamefont {Schalk}}, \bibinfo {author} {\bibfnamefont {P.}~\bibnamefont {Weinbrenner}}, \bibinfo {author} {\bibfnamefont {A.}~\bibnamefont {Angerer}}, \bibinfo {author} {\bibfnamefont {J.}~\bibnamefont {Majer}},\ and\ \bibinfo {author} {\bibfnamefont {F.}~\bibnamefont {Reinhard}},\ }\bibfield  {title} {\bibinfo {title} {Dispersive readout of room-temperature ensemble spin sensors},\ }\href {https://doi.org/10.1088/2058-9565/abfaaf} {\bibfield  {journal} {\bibinfo  {journal} {Quantum Sci. Technol.}\ }\textbf {\bibinfo {volume} {6}},\ \bibinfo {pages} {03LT01} (\bibinfo {year} {2021})}\BibitemShut {NoStop}%
\bibitem [{\citenamefont {Eisenach}\ \emph {et~al.}(2021)\citenamefont {Eisenach}, \citenamefont {Barry}, \citenamefont {O'Keeffe}, \citenamefont {Schloss}, \citenamefont {Steinecker}, \citenamefont {Englund},\ and\ \citenamefont {Braje}}]{eisenach_cavityenhanced_2021}%
  \BibitemOpen
  \bibfield  {author} {\bibinfo {author} {\bibfnamefont {E.~R.}\ \bibnamefont {Eisenach}}, \bibinfo {author} {\bibfnamefont {J.~F.}\ \bibnamefont {Barry}}, \bibinfo {author} {\bibfnamefont {M.~F.}\ \bibnamefont {O'Keeffe}}, \bibinfo {author} {\bibfnamefont {J.~M.}\ \bibnamefont {Schloss}}, \bibinfo {author} {\bibfnamefont {M.~H.}\ \bibnamefont {Steinecker}}, \bibinfo {author} {\bibfnamefont {D.~R.}\ \bibnamefont {Englund}},\ and\ \bibinfo {author} {\bibfnamefont {D.~A.}\ \bibnamefont {Braje}},\ }\bibfield  {title} {\bibinfo {title} {Cavity-enhanced microwave readout of a solid-state spin sensor},\ }\href {https://doi.org/10.1038/s41467-021-21256-7} {\bibfield  {journal} {\bibinfo  {journal} {Nat. Commun.}\ }\textbf {\bibinfo {volume} {12}},\ \bibinfo {pages} {1357} (\bibinfo {year} {2021})}\BibitemShut {NoStop}%
\bibitem [{\citenamefont {Wang}\ \emph {et~al.}(2024)\citenamefont {Wang}, \citenamefont {Tiwari}, \citenamefont {Jacobs}, \citenamefont {Judy}, \citenamefont {Zhang}, \citenamefont {Englund},\ and\ \citenamefont {Trusheim}}]{wang_spinrefrigerated_2024}%
  \BibitemOpen
  \bibfield  {author} {\bibinfo {author} {\bibfnamefont {H.}~\bibnamefont {Wang}}, \bibinfo {author} {\bibfnamefont {K.~L.}\ \bibnamefont {Tiwari}}, \bibinfo {author} {\bibfnamefont {K.}~\bibnamefont {Jacobs}}, \bibinfo {author} {\bibfnamefont {M.}~\bibnamefont {Judy}}, \bibinfo {author} {\bibfnamefont {X.}~\bibnamefont {Zhang}}, \bibinfo {author} {\bibfnamefont {D.~R.}\ \bibnamefont {Englund}},\ and\ \bibinfo {author} {\bibfnamefont {M.~E.}\ \bibnamefont {Trusheim}},\ }\bibfield  {title} {\bibinfo {title} {A spin-refrigerated cavity quantum electrodynamic sensor},\ }\href {https://doi.org/10.1038/s41467-024-54333-8} {\bibfield  {journal} {\bibinfo  {journal} {Nat. Commun.}\ }\textbf {\bibinfo {volume} {15}},\ \bibinfo {pages} {10320} (\bibinfo {year} {2024})},\ \Eprint {https://arxiv.org/abs/2404.10628} {arXiv:2404.10628} \BibitemShut {NoStop}%
\bibitem [{\citenamefont {Wu}\ \emph {et~al.}(2022)\citenamefont {Wu}, \citenamefont {Yang}, \citenamefont {Oxborrow}, \citenamefont {Jiang}, \citenamefont {Zhao}, \citenamefont {Budker}, \citenamefont {Zhang},\ and\ \citenamefont {Du}}]{wu_enhanced_2022}%
  \BibitemOpen
  \bibfield  {author} {\bibinfo {author} {\bibfnamefont {H.}~\bibnamefont {Wu}}, \bibinfo {author} {\bibfnamefont {S.}~\bibnamefont {Yang}}, \bibinfo {author} {\bibfnamefont {M.}~\bibnamefont {Oxborrow}}, \bibinfo {author} {\bibfnamefont {M.}~\bibnamefont {Jiang}}, \bibinfo {author} {\bibfnamefont {Q.}~\bibnamefont {Zhao}}, \bibinfo {author} {\bibfnamefont {D.}~\bibnamefont {Budker}}, \bibinfo {author} {\bibfnamefont {B.}~\bibnamefont {Zhang}},\ and\ \bibinfo {author} {\bibfnamefont {J.}~\bibnamefont {Du}},\ }\bibfield  {title} {\bibinfo {title} {Enhanced quantum sensing with room-temperature solid-state masers},\ }\href {https://doi.org/10.1126/sciadv.ade1613} {\bibfield  {journal} {\bibinfo  {journal} {Sci. Adv.}\ }\textbf {\bibinfo {volume} {8}},\ \bibinfo {pages} {eade1613} (\bibinfo {year} {2022})}\BibitemShut {NoStop}%
\bibitem [{\citenamefont {Han}\ \emph {et~al.}(2026)\citenamefont {Han}, \citenamefont {Wu}, \citenamefont {Wu}, \citenamefont {Oxborrow}, \citenamefont {Li}, \citenamefont {Wang}, \citenamefont {Zheng}, \citenamefont {Budker}, \citenamefont {Zhang},\ and\ \citenamefont {Zhang}}]{han_ultrasensitive_2026}%
  \BibitemOpen
  \bibfield  {author} {\bibinfo {author} {\bibfnamefont {Y.}~\bibnamefont {Han}}, \bibinfo {author} {\bibfnamefont {H.}~\bibnamefont {Wu}}, \bibinfo {author} {\bibfnamefont {H.}~\bibnamefont {Wu}}, \bibinfo {author} {\bibfnamefont {M.}~\bibnamefont {Oxborrow}}, \bibinfo {author} {\bibfnamefont {W.}~\bibnamefont {Li}}, \bibinfo {author} {\bibfnamefont {Y.}~\bibnamefont {Wang}}, \bibinfo {author} {\bibfnamefont {D.}~\bibnamefont {Zheng}}, \bibinfo {author} {\bibfnamefont {D.}~\bibnamefont {Budker}}, \bibinfo {author} {\bibfnamefont {B.}~\bibnamefont {Zhang}},\ and\ \bibinfo {author} {\bibfnamefont {J.}~\bibnamefont {Zhang}},\ }\bibfield  {title} {\bibinfo {title} {Ultra-sensitive microwave magnetometry with organic-molecular sensors},\ }\href {https://doi.org/10.1063/5.0271776} {\bibfield  {journal} {\bibinfo  {journal} {Appl. Phys. Rev.}\ }\textbf {\bibinfo {volume} {13}},\ \bibinfo {pages} {011403} (\bibinfo {year} {2026})}\BibitemShut {NoStop}%
\bibitem [{\citenamefont {Wang}\ \emph {et~al.}(2026)\citenamefont {Wang}, \citenamefont {Jacobs}, \citenamefont {Fahey}, \citenamefont {Hu}, \citenamefont {Englund},\ and\ \citenamefont {Trusheim}}]{wang_exceptional_2026}%
  \BibitemOpen
  \bibfield  {author} {\bibinfo {author} {\bibfnamefont {H.}~\bibnamefont {Wang}}, \bibinfo {author} {\bibfnamefont {K.}~\bibnamefont {Jacobs}}, \bibinfo {author} {\bibfnamefont {D.}~\bibnamefont {Fahey}}, \bibinfo {author} {\bibfnamefont {Y.}~\bibnamefont {Hu}}, \bibinfo {author} {\bibfnamefont {D.~R.}\ \bibnamefont {Englund}},\ and\ \bibinfo {author} {\bibfnamefont {M.~E.}\ \bibnamefont {Trusheim}},\ }\bibfield  {title} {\bibinfo {title} {Exceptional sensitivity near the bistable transition point of a hybrid quantum system},\ }\href {https://doi.org/10.1038/s41567-026-03217-3} {\bibfield  {journal} {\bibinfo  {journal} {Nat. Phys.}\ }\textbf {\bibinfo {volume} {22}},\ \bibinfo {pages} {577} (\bibinfo {year} {2026})}\BibitemShut {NoStop}%
\bibitem [{\citenamefont {Kapitanova}\ \emph {et~al.}(2018)\citenamefont {Kapitanova}, \citenamefont {Soshenko}, \citenamefont {Vorobyov}, \citenamefont {Dobrykh}, \citenamefont {Bolshedvorskii}, \citenamefont {Sorokin},\ and\ \citenamefont {Akimov}}]{kapitanova_3d_2018}%
  \BibitemOpen
  \bibfield  {author} {\bibinfo {author} {\bibfnamefont {P.}~\bibnamefont {Kapitanova}}, \bibinfo {author} {\bibfnamefont {V.~V.}\ \bibnamefont {Soshenko}}, \bibinfo {author} {\bibfnamefont {V.~V.}\ \bibnamefont {Vorobyov}}, \bibinfo {author} {\bibfnamefont {D.}~\bibnamefont {Dobrykh}}, \bibinfo {author} {\bibfnamefont {S.~V.}\ \bibnamefont {Bolshedvorskii}}, \bibinfo {author} {\bibfnamefont {V.~N.}\ \bibnamefont {Sorokin}},\ and\ \bibinfo {author} {\bibfnamefont {A.~V.}\ \bibnamefont {Akimov}},\ }\bibfield  {title} {\bibinfo {title} {{{3D Uniform Manipulation}} of {{NV Centers}} in {{Diamond Using}} a {{Dielectric Resonator Antenna}}},\ }\href {https://doi.org/10.1134/S0021364018210014} {\bibfield  {journal} {\bibinfo  {journal} {JETP Lett.}\ }\textbf {\bibinfo {volume} {108}},\ \bibinfo {pages} {588} (\bibinfo {year} {2018})}\BibitemShut {NoStop}%
\bibitem [{\citenamefont {Vallabhapurapu}\ \emph {et~al.}(2021)\citenamefont {Vallabhapurapu}, \citenamefont {{Slack-Smith}}, \citenamefont {Sewani}, \citenamefont {Adambukulam}, \citenamefont {Morello}, \citenamefont {Pla},\ and\ \citenamefont {Laucht}}]{vallabhapurapu_fast_2021}%
  \BibitemOpen
  \bibfield  {author} {\bibinfo {author} {\bibfnamefont {H.~H.}\ \bibnamefont {Vallabhapurapu}}, \bibinfo {author} {\bibfnamefont {J.~P.}\ \bibnamefont {{Slack-Smith}}}, \bibinfo {author} {\bibfnamefont {V.~K.}\ \bibnamefont {Sewani}}, \bibinfo {author} {\bibfnamefont {C.}~\bibnamefont {Adambukulam}}, \bibinfo {author} {\bibfnamefont {A.}~\bibnamefont {Morello}}, \bibinfo {author} {\bibfnamefont {J.~J.}\ \bibnamefont {Pla}},\ and\ \bibinfo {author} {\bibfnamefont {A.}~\bibnamefont {Laucht}},\ }\bibfield  {title} {\bibinfo {title} {Fast coherent control of a nitrogen-vacancy-center spin ensemble using a {$\mathrm{KTaO}_3$} dielectric resonator at cryogenic temperatures},\ }\href {https://doi.org/10.1103/PhysRevApplied.16.044051} {\bibfield  {journal} {\bibinfo  {journal} {Phys. Rev. Appl.}\ }\textbf {\bibinfo {volume} {16}},\ \bibinfo {pages} {044051} (\bibinfo {year} {2021})}\BibitemShut {NoStop}%
\bibitem [{\citenamefont {Alsid}\ \emph {et~al.}(2023)\citenamefont {Alsid}, \citenamefont {Schloss}, \citenamefont {Steinecker}, \citenamefont {Barry}, \citenamefont {Maccabe}, \citenamefont {Wang}, \citenamefont {Cappellaro},\ and\ \citenamefont {Braje}}]{alsid_solidstate_2023}%
  \BibitemOpen
  \bibfield  {author} {\bibinfo {author} {\bibfnamefont {S.~T.}\ \bibnamefont {Alsid}}, \bibinfo {author} {\bibfnamefont {J.~M.}\ \bibnamefont {Schloss}}, \bibinfo {author} {\bibfnamefont {M.~H.}\ \bibnamefont {Steinecker}}, \bibinfo {author} {\bibfnamefont {J.~F.}\ \bibnamefont {Barry}}, \bibinfo {author} {\bibfnamefont {A.~C.}\ \bibnamefont {Maccabe}}, \bibinfo {author} {\bibfnamefont {G.}~\bibnamefont {Wang}}, \bibinfo {author} {\bibfnamefont {P.}~\bibnamefont {Cappellaro}},\ and\ \bibinfo {author} {\bibfnamefont {D.~A.}\ \bibnamefont {Braje}},\ }\bibfield  {title} {\bibinfo {title} {Solid-{{State Microwave Magnetometer}} with {{Picotesla-Level Sensitivity}}},\ }\href {https://doi.org/10.1103/PhysRevApplied.19.054095} {\bibfield  {journal} {\bibinfo  {journal} {Phys. Rev. Appl.}\ }\textbf {\bibinfo {volume} {19}},\ \bibinfo {pages} {054095} (\bibinfo {year} {2023})}\BibitemShut {NoStop}%
\bibitem [{\citenamefont {Borneman}\ and\ \citenamefont {Cory}(2012)}]{borneman_bandwidthlimited_2012}%
  \BibitemOpen
  \bibfield  {author} {\bibinfo {author} {\bibfnamefont {T.~W.}\ \bibnamefont {Borneman}}\ and\ \bibinfo {author} {\bibfnamefont {D.~G.}\ \bibnamefont {Cory}},\ }\bibfield  {title} {\bibinfo {title} {Bandwidth-limited control and ringdown suppression in high-{$Q$} resonators},\ }\href {https://doi.org/10.1016/j.jmr.2012.10.011} {\bibfield  {journal} {\bibinfo  {journal} {J. Magn. Reson.}\ }\textbf {\bibinfo {volume} {225}},\ \bibinfo {pages} {120} (\bibinfo {year} {2012})}\BibitemShut {NoStop}%
\bibitem [{\citenamefont {Probst}\ \emph {et~al.}(2019)\citenamefont {Probst}, \citenamefont {Ranjan}, \citenamefont {Ansel}, \citenamefont {Heeres}, \citenamefont {Albanese}, \citenamefont {Albertinale}, \citenamefont {Vion}, \citenamefont {Esteve}, \citenamefont {Glaser}, \citenamefont {Sugny},\ and\ \citenamefont {Bertet}}]{probst_shaped_2019}%
  \BibitemOpen
  \bibfield  {author} {\bibinfo {author} {\bibfnamefont {S.}~\bibnamefont {Probst}}, \bibinfo {author} {\bibfnamefont {V.}~\bibnamefont {Ranjan}}, \bibinfo {author} {\bibfnamefont {Q.}~\bibnamefont {Ansel}}, \bibinfo {author} {\bibfnamefont {R.}~\bibnamefont {Heeres}}, \bibinfo {author} {\bibfnamefont {B.}~\bibnamefont {Albanese}}, \bibinfo {author} {\bibfnamefont {E.}~\bibnamefont {Albertinale}}, \bibinfo {author} {\bibfnamefont {D.}~\bibnamefont {Vion}}, \bibinfo {author} {\bibfnamefont {D.}~\bibnamefont {Esteve}}, \bibinfo {author} {\bibfnamefont {S.~J.}\ \bibnamefont {Glaser}}, \bibinfo {author} {\bibfnamefont {D.}~\bibnamefont {Sugny}},\ and\ \bibinfo {author} {\bibfnamefont {P.}~\bibnamefont {Bertet}},\ }\bibfield  {title} {\bibinfo {title} {Shaped pulses for transient compensation in quantum-limited electron spin resonance spectroscopy},\ }\href {https://doi.org/10.1016/j.jmr.2019.04.008} {\bibfield  {journal} {\bibinfo  {journal} {J. Magn. Reson.}\ }\textbf {\bibinfo {volume} {303}},\ \bibinfo {pages}
  {42} (\bibinfo {year} {2019})}\BibitemShut {NoStop}%
\bibitem [{\citenamefont {Ansel}\ \emph {et~al.}(2018)\citenamefont {Ansel}, \citenamefont {Probst}, \citenamefont {Bertet}, \citenamefont {Glaser},\ and\ \citenamefont {Sugny}}]{ansel_optimal_2018}%
  \BibitemOpen
  \bibfield  {author} {\bibinfo {author} {\bibfnamefont {Q.}~\bibnamefont {Ansel}}, \bibinfo {author} {\bibfnamefont {S.}~\bibnamefont {Probst}}, \bibinfo {author} {\bibfnamefont {P.}~\bibnamefont {Bertet}}, \bibinfo {author} {\bibfnamefont {S.~J.}\ \bibnamefont {Glaser}},\ and\ \bibinfo {author} {\bibfnamefont {D.}~\bibnamefont {Sugny}},\ }\bibfield  {title} {\bibinfo {title} {Optimal control of an inhomogeneous spin ensemble coupled to a cavity},\ }\href {https://doi.org/10.1103/PhysRevA.98.023425} {\bibfield  {journal} {\bibinfo  {journal} {Phys. Rev. A}\ }\textbf {\bibinfo {volume} {98}},\ \bibinfo {pages} {023425} (\bibinfo {year} {2018})}\BibitemShut {NoStop}%
\bibitem [{\citenamefont {{Iriarte-Zendoia}}\ \emph {et~al.}(2025)\citenamefont {{Iriarte-Zendoia}}, \citenamefont {{Munuera-Javaloy}},\ and\ \citenamefont {Casanova}}]{iriarte-zendoia_robust_2025}%
  \BibitemOpen
  \bibfield  {author} {\bibinfo {author} {\bibfnamefont {I.}~\bibnamefont {{Iriarte-Zendoia}}}, \bibinfo {author} {\bibfnamefont {C.}~\bibnamefont {{Munuera-Javaloy}}},\ and\ \bibinfo {author} {\bibfnamefont {J.}~\bibnamefont {Casanova}},\ }\bibfield  {title} {\bibinfo {title} {Robust microwave cavity control for {{NV}} ensemble manipulation},\ }\href {https://doi.org/10.1103/PhysRevResearch.7.013315} {\bibfield  {journal} {\bibinfo  {journal} {Phys. Rev. Res.}\ }\textbf {\bibinfo {volume} {7}},\ \bibinfo {pages} {013315} (\bibinfo {year} {2025})}\BibitemShut {NoStop}%
\bibitem [{\citenamefont {Rosenblum}\ \emph {et~al.}(2015)\citenamefont {Rosenblum}, \citenamefont {Lovsky}, \citenamefont {Arazi}, \citenamefont {Vollmer},\ and\ \citenamefont {Dayan}}]{rosenblum_cavity_2015}%
  \BibitemOpen
  \bibfield  {author} {\bibinfo {author} {\bibfnamefont {S.}~\bibnamefont {Rosenblum}}, \bibinfo {author} {\bibfnamefont {Y.}~\bibnamefont {Lovsky}}, \bibinfo {author} {\bibfnamefont {L.}~\bibnamefont {Arazi}}, \bibinfo {author} {\bibfnamefont {F.}~\bibnamefont {Vollmer}},\ and\ \bibinfo {author} {\bibfnamefont {B.}~\bibnamefont {Dayan}},\ }\bibfield  {title} {\bibinfo {title} {Cavity ring-up spectroscopy for ultrafast sensing with optical microresonators},\ }\href {https://doi.org/10.1038/ncomms7788} {\bibfield  {journal} {\bibinfo  {journal} {Nat. Commun.}\ }\textbf {\bibinfo {volume} {6}},\ \bibinfo {pages} {6788} (\bibinfo {year} {2015})}\BibitemShut {NoStop}%
\bibitem [{\citenamefont {Gy{\"u}re}\ \emph {et~al.}(2015)\citenamefont {Gy{\"u}re}, \citenamefont {M{\'a}rkus}, \citenamefont {Bern{\'a}th}, \citenamefont {Mur{\'a}nyi},\ and\ \citenamefont {Simon}}]{gyure_time_2015}%
  \BibitemOpen
  \bibfield  {author} {\bibinfo {author} {\bibfnamefont {B.}~\bibnamefont {Gy{\"u}re}}, \bibinfo {author} {\bibfnamefont {B.}~\bibnamefont {M{\'a}rkus}}, \bibinfo {author} {\bibfnamefont {B.}~\bibnamefont {Bern{\'a}th}}, \bibinfo {author} {\bibfnamefont {F.}~\bibnamefont {Mur{\'a}nyi}},\ and\ \bibinfo {author} {\bibfnamefont {F.}~\bibnamefont {Simon}},\ }\bibfield  {title} {\bibinfo {title} {A time domain based method for the accurate measurement of {Q}-factor and resonance frequency of microwave resonators},\ }\href {https://doi.org/10.1063/1.4929865} {\bibfield  {journal} {\bibinfo  {journal} {Rev. Sci. Instrum.}\ }\textbf {\bibinfo {volume} {86}},\ \bibinfo {pages} {094702} (\bibinfo {year} {2015})}\BibitemShut {NoStop}%
\bibitem [{\citenamefont {Lenz}\ \emph {et~al.}(2021)\citenamefont {Lenz}, \citenamefont {K{\"o}nig}, \citenamefont {Hunger},\ and\ \citenamefont {van Slageren}}]{lenz2021room}%
  \BibitemOpen
  \bibfield  {author} {\bibinfo {author} {\bibfnamefont {S.}~\bibnamefont {Lenz}}, \bibinfo {author} {\bibfnamefont {D.}~\bibnamefont {K{\"o}nig}}, \bibinfo {author} {\bibfnamefont {D.}~\bibnamefont {Hunger}},\ and\ \bibinfo {author} {\bibfnamefont {J.}~\bibnamefont {van Slageren}},\ }\bibfield  {title} {\bibinfo {title} {Room-temperature quantum memories based on molecular electron spin ensembles},\ }\href {https://doi.org/10.1002/adma.202101673} {\bibfield  {journal} {\bibinfo  {journal} {Adv. Mater.}\ }\textbf {\bibinfo {volume} {33}},\ \bibinfo {pages} {2101673} (\bibinfo {year} {2021})}\BibitemShut {NoStop}%
\bibitem [{SOM()}]{SOM}%
  \BibitemOpen
  \href@noop {} {}\bibinfo {note} {See Supplemental Material at [URL] for additional details on theoretical derivations of spin dynamics under cavity drives, experimental details, and numerical simulations.}\BibitemShut {Stop}%
\bibitem [{\citenamefont {Leroux}\ \emph {et~al.}(2010)\citenamefont {Leroux}, \citenamefont {Schleier-Smith},\ and\ \citenamefont {Vuleti{\'c}}}]{leroux_implementation_2010}%
  \BibitemOpen
  \bibfield  {author} {\bibinfo {author} {\bibfnamefont {I.~D.}\ \bibnamefont {Leroux}}, \bibinfo {author} {\bibfnamefont {M.~H.}\ \bibnamefont {Schleier-Smith}},\ and\ \bibinfo {author} {\bibfnamefont {V.}~\bibnamefont {Vuleti{\'c}}},\ }\bibfield  {title} {\bibinfo {title} {Implementation of cavity squeezing of a collective atomic spin},\ }\href {https://doi.org/10.1103/PhysRevLett.104.073602} {\bibfield  {journal} {\bibinfo  {journal} {Phys. Rev. Lett.}\ }\textbf {\bibinfo {volume} {104}},\ \bibinfo {pages} {073602} (\bibinfo {year} {2010})}\BibitemShut {NoStop}%
\bibitem [{\citenamefont {Fukumori}\ \emph {et~al.}(2026)\citenamefont {Fukumori}, \citenamefont {Luo}, \citenamefont {Tiranov}, \citenamefont {Waszkowska}, \citenamefont {Goldner},\ and\ \citenamefont {Faraon}}]{fukumori_cavitymediated_2026a}%
  \BibitemOpen
  \bibfield  {author} {\bibinfo {author} {\bibfnamefont {R.}~\bibnamefont {Fukumori}}, \bibinfo {author} {\bibfnamefont {C.}~\bibnamefont {Luo}}, \bibinfo {author} {\bibfnamefont {A.}~\bibnamefont {Tiranov}}, \bibinfo {author} {\bibfnamefont {K.}~\bibnamefont {Waszkowska}}, \bibinfo {author} {\bibfnamefont {P.}~\bibnamefont {Goldner}},\ and\ \bibinfo {author} {\bibfnamefont {A.}~\bibnamefont {Faraon}},\ }\href@noop {} {\bibinfo {title} {Cavity-mediated coherence protection and one-axis twisting for spins in solids}} (\bibinfo {year} {2026}),\ \Eprint {https://arxiv.org/abs/2604.26909} {arXiv:2604.26909 [quant-ph]} \BibitemShut {NoStop}%
\bibitem [{\citenamefont {Wang}\ \emph {et~al.}(2020)\citenamefont {Wang}, \citenamefont {Liu},\ and\ \citenamefont {Cappellaro}}]{wang_coherence_2020}%
  \BibitemOpen
  \bibfield  {author} {\bibinfo {author} {\bibfnamefont {G.}~\bibnamefont {Wang}}, \bibinfo {author} {\bibfnamefont {Y.-X.}\ \bibnamefont {Liu}},\ and\ \bibinfo {author} {\bibfnamefont {P.}~\bibnamefont {Cappellaro}},\ }\bibfield  {title} {\bibinfo {title} {Coherence protection and decay mechanism in qubit ensembles under concatenated continuous driving},\ }\href {https://doi.org/10.1088/1367-2630/abd2e5} {\bibfield  {journal} {\bibinfo  {journal} {New Journal of Physics}\ }\textbf {\bibinfo {volume} {22}},\ \bibinfo {pages} {123045} (\bibinfo {year} {2020})}\BibitemShut {NoStop}%
\end{thebibliography}%

\onecolumngrid
\clearpage
\begin{center}
\textbf{\large Supplemental Material: Spin Interferometry Generated by Cavity Memory}
\end{center}
\setcounter{section}{0}
\setcounter{equation}{0}
\setcounter{figure}{0}
\setcounter{table}{0}
\setcounter{page}{1}
\makeatletter
\renewcommand{\theequation}{S\arabic{equation}}
\renewcommand{\thesection}{S\arabic{section}}
\renewcommand{\thefigure}{S\arabic{figure}}


\tableofcontents

\section{Theory}

\subsection{Ramsey interferometry}
\label{sec:ramsey_limit}

Here we show that the spin dynamics induced by the ring-up and ring-down of a driven cavity reduce to an effective Ramsey sequence. We first discuss the ideal Ramsey limit in the cavity rotating frame to analyze the transient-induced spin rotation. We then move to the spin rotating frame and show that the characteristic oscillation at the drive-spin detuning persists much more generally, even when the cavity transients do not realize ideal $\pi/2$ rotations. Finally, we analyze the coherence time of the Ramsey fringes when the cavity transient is comparable with the evolution time.

\subsubsection{Ideal Ramsey interferometry derived in the cavity rotating frame}
We consider a general two-level spin transition at frequency $\omega_s$, coupled to a cavity at $\omega_c$ and driven at $\omega_d$. We define the detunings as 
$\Delta_{\mathrm{dc}}\equiv\omega_d-\omega_c,
\Delta_{\mathrm{sc}}\equiv\omega_s-\omega_c,
\Delta_{\mathrm{ds}}\equiv\omega_d-\omega_s
=\Delta_{\mathrm{dc}}-\Delta_{\mathrm{sc}}$.
Here $\Delta_{\mathrm{sc}}$ is the spin-cavity detuning and $\Delta_{\mathrm{ds}}$ is the drive-spin detuning. As in the main text, $t$ denotes the applied pulse duration, $t_1$ is the running time measured from turn-on, and $t_2\geq0$ is the elapsed time after switch-off. 
The laboratory-frame field amplitude is denoted by $\alpha$, as in the main text. A superscript identifies a rotating-frame amplitude: $\alpha^{(c)}(t_1)=e^{i\omega_c t_1}\alpha(t_1)$ and $\alpha^{(d)}(t_1)=e^{i\omega_d t_1}\alpha(t_1)$. 

For convenience in analyzing the cavity dynamics, we first work in a frame rotating at the cavity frequency $\omega_c$. For a square microwave pulse switched on at $t_1=0$, the intracavity field amplitude $\alpha^{(c)}(t_1)$ obeys
\begin{equation}
    \dot{\alpha}^{(c)}(t_1)
    =
    -\frac{\kappa}{2}\alpha^{(c)}(t_1)
    +
    \eta e^{-i\Delta_{\mathrm{dc}}t_1},
    \label{eq:cavity_eom_cavity_frame}
\end{equation}
where $\kappa$ is the cavity energy decay rate, corresponding to the full width at half maximum linewidth of the cavity resonance, and $\eta$ denotes the drive amplitude. For an initially empty cavity, $\alpha^{(c)}(0)=0$, the solution during the drive pulse is
\begin{equation}
    \alpha_{\rm on}^{(c)}(t_1)
    =
    \bar{\alpha}
    \left(
        e^{-i\Delta_{\mathrm{dc}}t_1}
        -
        e^{-\kappa t_1/2}
    \right),
    \qquad
    \bar{\alpha}
    =
    \frac{\eta}{\kappa/2-i\Delta_{\mathrm{dc}}}.
    \label{eq:alpha_ringup}
\end{equation}
Equation~(\ref{eq:alpha_ringup}) separates the intracavity field into two components. The first term is the forced response at the drive frequency and therefore carries the phase $e^{-i\Delta_{\mathrm{dc}}t_1}$ in the cavity frame. The second term is the freely evolving cavity transient, which is stationary in this frame and decays as $e^{-\kappa t_1/2}$.

We absorb the constant phase of $\bar{\alpha}$ into the definition of the transverse spin axis and define $\Omega_d=\gamma_e|\bar{\alpha}|$ as the steady-state spin Rabi frequency associated with the magnetic field strength $|\bar{\alpha}|$, where $\gamma_e$ is the gyromagnetic ratio. Within the rotating-wave approximation, the spin Hamiltonian in the cavity rotating frame is
\begin{equation}
    H_{\rm on}(t_1)=
    -\frac{\Delta_{\mathrm{sc}}}{2}\sigma_z
    +
    \frac{\Omega_d}{2}
    \left[
        \cos(\Delta_{\mathrm{dc}}t_1)\sigma_x
        -
        \sin(\Delta_{\mathrm{dc}}t_1)\sigma_y
    \right]
    -
    \frac{\Omega_d}{2}
    e^{-\kappa t_1/2}\sigma_x.
    \label{eq:H_ringup_vector}
\end{equation}
The second term describes the continuously applied microwave drive, whereas the last term originates from the cavity ring-up transient.

We now consider the regime in which the continuous drive is far detuned from the spin transition, $|\Delta_{\mathrm{ds}}|\gg\Omega_d$. Its leading-order direct action on the spin averages out, leaving only higher-order corrections such as an ac-Stark shift of order $\Omega_d^2/|\Delta_{\mathrm{ds}}|$. The spin Hamiltonian during the ring-up transient therefore reduces approximately to
\begin{equation}
    H_{\rm on}(t_1)
    \simeq
    -\frac{\Delta_{\mathrm{sc}}}{2}\sigma_z
    -
    \frac{\Omega_d}{2}e^{-\kappa t_1/2}\sigma_x.
    \label{eq:H_ringup_approx}
\end{equation}
When the spin lies sufficiently close to the cavity resonance, $|\Delta_{\mathrm{sc}}|\ll\kappa$, the spin precession generated by $\Delta_{\mathrm{sc}}$ is negligible over the duration of the cavity transient, and
$H_{\rm on}(t_1)  \simeq -\frac{\Omega_d}{2}e^{-\kappa t_1/2}\sigma_x$.
The ring-up transient therefore produces a rotation angle $\theta_{\rm on}=\int_0^\infty\Omega_d e^{-\kappa t_1/2}dt_1 =2\Omega_d/\kappa$. In particular, it realizes an effective $\pi/2$ rotation when $\Omega_d=\frac{\pi\kappa}{4}$.

After several cavity lifetimes, the transient term has decayed, leaving $H_{\rm mid}\simeq-\Delta_{\mathrm{sc}}\sigma_z/2$, the spin approximately undergoes free precession at $-\Delta_{\mathrm{sc}}$. We now switch off the microwave drive at time $t_1=t$. Immediately before switch-off, the cavity field is   $\alpha^{(c)}(t) =\bar{\alpha} \left(  e^{-i\Delta_{\mathrm{dc}}t}  - e^{-\kappa t/2}    \right)$.
After switch-off, defining $t_2=t_1-t$, the cavity evolves freely as
\begin{equation}
    \alpha_{\rm off}^{(c)}(t+t_2)
    =
    \bar{\alpha}
    \left(
        e^{-i\Delta_{\mathrm{dc}}t}
        -
        e^{-\kappa t/2}
    \right)
    e^{-\kappa t_2/2}.
    \label{eq:alpha_ringdown_exact}
\end{equation}
For a pulse duration long compared with the cavity transient time, $t\gg2/\kappa$, the residual contribution from the initial ring-up transient is negligible, giving
$\alpha_{\rm off}^{(c)}(t+t_2)\simeq\bar{\alpha}e^{-i\Delta_{\mathrm{dc}}t}e^{-\kappa t_2/2}$.
The corresponding spin Hamiltonian is
\begin{equation}
    H_{\rm off}(t+t_2)=
    -\frac{\Delta_{\mathrm{sc}}}{2}\sigma_z
    +
    \frac{\Omega_d}{2}e^{-\kappa t_2/2}
    \left[
        \cos(\Delta_{\mathrm{dc}}t)\sigma_x
        -
        \sin(\Delta_{\mathrm{dc}}t)\sigma_y
    \right].
    \label{eq:H_ringdown_vector}
\end{equation}
Under the additional condition $|\Delta_{\mathrm{sc}}|\ll\kappa$, the $\sigma_z$ term is again negligible over the duration of the transient. The ring-down therefore realizes a second transverse rotation with the same angle $2\Omega_d/\kappa$, while its rotation axis is shifted in the transverse plane by $-\Delta_{\mathrm{dc}}t$.

For $\Omega_d=\pi\kappa/4$, the two cavity transients thus realize approximately two $\pi/2$ rotations. The spin precesses between them by $-\Delta_{\mathrm{sc}}t$, whereas the analysis axis rotates by $-\Delta_{\mathrm{dc}}t$. Defining the relative interferometric phase as the spin-precession angle minus the analysis-axis angle gives
$\Phi(t) =\Delta_{\mathrm{dc}}t-\Delta_{\mathrm{sc}}t =  \Delta_{\mathrm{ds}}t$, 
and the population exhibits the familiar Ramsey oscillation at $|\Delta_{\mathrm{ds}}|=|\omega_d-\omega_s|$. Thus, in this ideal limit, a single square microwave pulse is converted by the finite cavity response into two approximately resonant $\pi/2$ rotations separated by an effectively free spin-evolution interval.

\subsubsection{General Ramsey interferometry derived in the spin rotating frame}
\label{sec:general_transient_interference}

The ideal Ramsey picture discussed above assumes $|\Delta_{\mathrm{sc}}|\ll\kappa$, such that the cavity ring-up and ring-down transients act approximately as fixed-axis transverse rotations. However, this condition is not required for the appearance of the detuning-locked oscillation. We now show that the same interference at $\Delta_{\mathrm{ds}}=\omega_d-\omega_s$ persists for arbitrary spin-cavity detuning and arbitrary transient rotation angles.

For this purpose, it is convenient to move to the spin rotating frame. During cavity ring-up, the transverse field seen by the spin is 
$\Omega_{\rm on}^{(s)}(t_1)=\Omega_d\left[    e^{-i\Delta_{\mathrm{ds}}t_1}    -    e^{-\kappa t_1/2}e^{i\Delta_{\mathrm{sc}}t_1}\right]$.
In the far-detuned regime $|\Delta_{\mathrm{ds}}|\gg\Omega_d$, the direct action of the continuously applied drive averages out to leading order. The remaining ring-up transient is therefore described by
$H_{\rm on}^{(s)}(t_1)\simeq-\frac{\Omega_d}{2}e^{-\kappa t_1/2}\left(e^{i\Delta_{\mathrm{sc}}t_1}\sigma_-+e^{-i\Delta_{\mathrm{sc}}t_1}\sigma_+\right)$. 
For finite $\Delta_{\mathrm{sc}}$, the corresponding transverse rotation axis changes continuously during the cavity decay. The resulting operation therefore need not be a simple fixed-axis rotation, and in particular does not need to be an ideal $\pi/2$ pulse. We denote the complete unitary generated by this transient as $U_{\rm on}$.

We next consider the ring-down transient. In the long-pulse limit $t\gg2/\kappa$, the cavity has reached its driven steady state before the microwave is switched off. In the spin rotating frame, the ring-down field is
$\Omega_{\rm off}^{(s)}(t+t_2)\simeq\Omega_d e^{-i\Delta_{\mathrm{ds}}t}e^{-\kappa t_2/2}
e^{i\Delta_{\mathrm{sc}}t_2}$, 
where $t_2=t_1-t$ denotes the time after switch-off. We first define a reference ring-down operation by removing the overall phase $e^{-i\Delta_{\mathrm{ds}}t}$. Its Hamiltonian is
$H_{\rm off}^{(0,s)}(t_2)=+\frac{\Omega_d}{2}e^{-\kappa t_2/2}\left(e^{i\Delta_{\mathrm{sc}}t_2}\sigma_-+e^{-i\Delta_{\mathrm{sc}}t_2}\sigma_+\right)$. 
Using $\sigma_z\sigma_\pm\sigma_z=-\sigma_\pm$, we obtain  $H_{\rm off}^{(0,s)}(t_2) =   \sigma_z H_{\rm on}^{(s)}(t_2)\sigma_z $.
Since the transformation by $\sigma_z$ is time independent, the same relation holds for the corresponding time-ordered evolution operators: $   U_{\rm off}^{(0)} = \sigma_z U_{\rm on}\sigma_z$.
This relation is independent of the detailed transient dynamics and, in particular, does not require $|\Delta_{\mathrm{sc}}|\ll\kappa$. Thus, even when the spin--cavity detuning causes the rotation axis to evolve during the transient, the ring-up and reference ring-down operations remain directly related by a $\pi$ rotation about the spin quantization axis. In the near-resonant limit $|\Delta_{\mathrm{sc}}|\ll\kappa$, the effective rotation axis becomes nearly transverse, and $U_{\rm off}^{(0)}\simeq U_{\rm on}^{\dagger}$.

For a finite pulse duration $t$, the additional factor $e^{-i\Delta_{\mathrm{ds}}t}$ in the ring-down field rotates the entire ring-down Hamiltonian in the transverse plane by $-\Delta_{\mathrm{ds}}t$. The actual ring-down unitary is therefore $U_{\rm off}(t)=R_z(-\Delta_{\mathrm{ds}}t) \sigma_z U_{\rm on}\sigma_z R_z^\dagger(-\Delta_{\mathrm{ds}}t)$, 
where $R_z(\phi)=e^{-i\phi\sigma_z/2}$. The detailed dynamics of $U_{\rm{on}}$ and $U_{\rm off}$ are the same, while the sign reversal and the scanned pulse duration enter through the relative phase. After the ring-up transient has decayed, the remaining continuous drive is far detuned from the spin and produces negligible direct evolution to leading order. The total evolution is consequently $U(t) \simeq  R_z(-\Delta_{\mathrm{ds}}t) \sigma_z U_{\rm on}\sigma_z R_z^\dagger(-\Delta_{\mathrm{ds}}t)
U_{\rm on}$. 
For an initial state $|0\rangle$ followed by a population measurement in the same basis, the leftmost $R_z(-\Delta_{\mathrm{ds}}t)$ and the adjacent $\sigma_z$ contribute only overall phases and therefore do not affect the measured probability. The signal can thus be written as
\begin{equation}
    P_0(t)
    =
    \left|
        \langle0|
        U_{\rm on}\sigma_z
        R_z^\dagger(-\Delta_{\mathrm{ds}}t)
        U_{\rm on}
        |0\rangle
    \right|^2,
    \label{eq:general_population_spinframe}
\end{equation}
up to an irrelevant overall phase.

The dependence on $t$ is now explicit. Let the state prepared by the ring-up transient be $U_{\rm on}|0\rangle=c_0|0\rangle+c_1|1\rangle$, where $|c_0|^2+|c_1|^2=1$. 
Using Eq.~(\ref{eq:general_population_spinframe}), the intermediate $z$ rotation gives $R_z^\dagger(-\Delta_{\mathrm{ds}}t)U_{\rm on}|0\rangle = c_0 e^{-i\Delta_{\mathrm{ds}}t/2}|0\rangle + c_1 e^{i\Delta_{\mathrm{ds}}t/2}|1\rangle$. After the subsequent $\sigma_z$ operation, this becomes $c_0 e^{-i\Delta_{\mathrm{ds}}t/2}|0\rangle -c_1 e^{i\Delta_{\mathrm{ds}}t/2}|1\rangle$. Applying the final $U_{\rm on}$ and projecting onto $|0\rangle$, the probability amplitude is therefore $\mathcal{A}_0(t)    =    c_0^2 e^{-i\Delta_{\mathrm{ds}}t/2}    +  |c_1|^2 e^{i\Delta_{\mathrm{ds}}t/2}$.
The measured population is consequently
\begin{align}
    P_0(t)
    &=
    |\mathcal{A}_0(t)|^2
    =
    |c_0|^4+|c_1|^4
    +
    2|c_0|^2|c_1|^2
    \cos\!\left(
        \Delta_{\mathrm{ds}}t-2\arg c_0
    \right).
    \label{eq:general_signal_c0c1}
\end{align}
The oscillation frequency is therefore 
$\omega_{\rm fringe}=|\Delta_{\mathrm{ds}}|
=|\omega_d-\omega_s|$.

This result shows that ideal $\pi/2$ rotations are not essential for the observed interference. The role of the ring-up transient is to generate transverse spin coherence, while the ring-down transient converts the orientation of this coherence back into a population difference. The detailed transient dynamics, including a finite spin--cavity detuning, determine the efficiency of these two processes and therefore the fringe contrast and phase offset. The oscillation frequency, however, is determined by the relative phase $\Delta_{\mathrm{ds}}t$ accumulated between the continuously applied drive and the spin. The ideal Ramsey limit corresponds to the special case in which the two transient operations maximize the preparation and readout of transverse coherence; the detuning-locked interference itself is considerably more general.

\subsubsection{Coherence analysis of the Ramsey decay envelope}
\label{sec:transient_coherence_decay}
The observed Ramsey fringes outlast the bare ensemble dephasing time $T_2^*\sim0.21\,\upmu\mathrm{s}$. During the nominal phase-accumulation interval, the residual cavity-frequency field continues to rotate the spins, so the fringe envelope need not follow free evolution. We examine this effect in the weak spin-backaction limit, treating the transient as a prescribed exponentially decaying drive. An externally synthesized field with the same complex waveform would give identical reduced spin dynamics.

For a cavity-resonant spin, $\omega_s=\omega_c$, choose the first transient as the transverse phase reference and write $H_{\rm tr}(t_1)=\Omega(t_1)\sigma_x/2$, with $\Omega(t_1)=\Omega_d e^{-\kappa t_1/2}$. We retain homogeneous Markovian dephasing at rate $\gamma_\perp$ and neglect longitudinal relaxation, $\gamma_1t\ll1$. Using $s_-=(s_x-is_y)/2$, $s_z=P_0-P_1$, $[\sigma_x,\sigma^-]=\sigma_z$, and $[\sigma_x,\sigma_z]=2(\sigma^--\sigma^+)$ gives
\begin{equation}
    \begin{aligned}
        \dot s_-&=i\langle[H_{\rm tr},\sigma^-]\rangle-\gamma_\perp s_-
        =-\gamma_\perp s_-+\frac{i\Omega(t_1)}{2}s_z,\\
        \dot s_z&=i\langle[H_{\rm tr},\sigma_z]\rangle
        =i\Omega(t_1)(s_--s_-^*).
    \end{aligned}
\end{equation}
Substituting $s_-=(s_x-is_y)/2$, so that $s_--s_-^*=-is_y$, and taking the real and imaginary parts of the first equation then yields
\begin{equation}
    \dot s_x=-\gamma_\perp s_x,\qquad
    \dot s_y=-\gamma_\perp s_y-\Omega(t_1)s_z,\qquad
    \dot s_z=\Omega(t_1)s_y.
    \label{eq:transient_decay_bloch}
\end{equation}

For a constant drive $\Omega$, the coupled $s_y,s_z$ components have eigenvalues $-\gamma_\perp/2\pm i\sqrt{\Omega^2-\gamma_\perp^2/4}$: underdamped oscillations decay as $e^{-\gamma_\perp t_1/2}$, while $s_x$ decays as $e^{-\gamma_\perp t_1}$. This difference reflects coherent exchange between population and transverse spin components.

For the exponentially decaying field, eliminating $s_y$ yields
\begin{equation}
    \ddot s_z+\left(\gamma_\perp+\frac{\kappa}{2}\right)\dot s_z
    +\Omega_d^2e^{-\kappa t_1}s_z=0.
    \label{eq:transient_population_ode}
\end{equation}
The corresponding coherence follows directly from
\begin{equation}
    s_y(t_1)=s_y(0)e^{-\gamma_\perp t_1}
    -\Omega_d\int_0^{t_1}
    e^{-\gamma_\perp(t_1-\tau)}e^{-\kappa\tau/2}s_z(\tau)\,d\tau,
    \qquad s_x(t_1)=s_x(0)e^{-\gamma_\perp t_1}.
    \label{eq:transient_coherence_integral}
\end{equation}
If $\gamma_\perp>\kappa/2$ and the population approaches a nonzero residual value $s_{z,\infty}$, the leading late-time terms are
\begin{align}
    s_y(t_1)&\simeq
    -\frac{\Omega_d s_{z,\infty}}{\gamma_\perp-\kappa/2}
    e^{-\kappa t_1/2},
    \label{eq:transient_coherence_tail}\\
    s_z(t_1)-s_{z,\infty}&\simeq
    \frac{\Omega_d^2s_{z,\infty}}
    {\kappa(\gamma_\perp-\kappa/2)}
    e^{-\kappa t_1}.
    \label{eq:transient_population_tail}
\end{align}
The remaining drive continually converts a residual population imbalance into coherence. Here $s_{z,\infty}$ is the population difference left by the finite-area transient with $T_1$ relaxation neglected, rather than a thermal-equilibrium value. The tails above require $\gamma_\perp>\kappa/2$ and $s_{z,\infty}\neq0$.

Immediately before switch-off, denote the spin components by $s_y(t),s_z(t)$, with initial conditions $s_x(0)=s_y(0)=0$ and $s_z(0)=1$. For matching switching-edge envelopes, the second transient has magnitude $\Omega_d e^{-\kappa t_2/2}$ and axis phase $\phi(t)=-\Delta_{\mathrm{dc}}t+\phi_0$, where $\phi_0$ includes the fixed edge-sign difference. In its axis-aligned coordinates, the coupled transverse component is $s_y(t)\cos\phi(t)$. The completed analysis operation gives $s_{z,f}=m_{zz}s_z+m_{zy}s_y\cos\phi$, and hence
\begin{equation}
    P_0(t)=\frac{1+m_{zz}s_z(t)}{2}
    +\frac{m_{zy}s_y(t)}{2}
    \cos\!\left(-\Delta_{\mathrm{dc}}t+\phi_0\right).
    \label{eq:transient_two_pulse_population}
\end{equation}
The coefficients $m_{zz},m_{zy}$ include dephasing during the analysis transient and are independent of $t$ within this approximation. The two terms separate the population background from the coherence converted into Ramsey fringes.

Substituting Eqs.~(\ref{eq:transient_coherence_tail}) and (\ref{eq:transient_population_tail}) into Eq.~(\ref{eq:transient_two_pulse_population}) gives
\begin{align}
    P_0(t)\simeq{}&
    \frac{1+m_{zz}s_{z,\infty}}{2}
    +\frac{m_{zz}\Omega_d^2s_{z,\infty}}
    {2\kappa(\gamma_\perp-\kappa/2)}e^{-\kappa t}-\frac{m_{zy}\Omega_ds_{z,\infty}}
    {2(\gamma_\perp-\kappa/2)}e^{-\kappa t/2}
    \cos\!\left(-\Delta_{\mathrm{dc}}t+\phi_0\right).
    \label{eq:transient_ramsey_decay}
\end{align}
Under the stated conditions, the background approaches its residual value on the timescale $1/\kappa$, while the fringe envelope follows the coherence tail on the timescale $2/\kappa$. For $\kappa/(2\pi)=0.25\,\mathrm{MHz}$, these times are $0.637\,\upmu\mathrm{s}$ and $1.27\,\upmu\mathrm{s}$, respectively.

The evolution of both transients can be obtained analytically. Set $x_0=2\Omega_d/\kappa$, $x(t_1)=x_0e^{-\kappa t_1/2}$, and $\nu=1/2+\gamma_\perp/\kappa$. Writing $s_z=x^\nu f(x)$ and using $d/dt_1=-(\kappa/2)x\,d/dx$ reduces Eq.~(\ref{eq:transient_population_ode}) to $x^2f^{\prime\prime}+xf^\prime+(x^2-\nu^2)f=0$, which is Bessel's equation. Since $s_y=-ds_z/dx$, the derivative identities for $x^\nu J_\nu(x)$ and $x^\nu Y_\nu(x)$ give the first row of the propagator below. With $J_\nu$ and $Y_\nu$ the Bessel functions of the first and second kind, define
\begin{equation}
    \mathsf F(x)=x^\nu
    \begin{pmatrix}
        -J_{\nu-1}(x)&-Y_{\nu-1}(x)\\
        J_\nu(x)&Y_\nu(x)
    \end{pmatrix},
    \qquad
    \begin{pmatrix}s_y(t)\\s_z(t)\end{pmatrix}
    =\mathsf F\!\left(x(t)\right)\mathsf F(x_0)^{-1}
    \begin{pmatrix}0\\1\end{pmatrix}.
    \label{eq:transient_bessel_propagator}
\end{equation}
Taking the end of the analysis transient to $t_2\rightarrow\infty$ gives
\begin{equation}
    m_{zz}=a_\nu J_{\nu-1}(x_0),\qquad
    m_{zy}=a_\nu J_\nu(x_0),\qquad
    a_\nu=2^{\nu-1}\Gamma_{\!E}(\nu)x_0^{1-\nu},
    \label{eq:transient_analysis_coefficients}
\end{equation}
where $\Gamma_{\!E}$ is the Euler gamma function. Equations~(\ref{eq:transient_two_pulse_population}), (\ref{eq:transient_bessel_propagator}), and(\ref{eq:transient_analysis_coefficients}) are exact for the prescribed resonant two-transient model. In the absence of dephasing, $m_{zz}=\cos x_0$ and $m_{zy}=\sin x_0$, recovering an analysis rotation with pulse area $x_0$.
An ideal $\pi/2$ analysis rotation has $m_{zz}=0$, $m_{zy}=1$ and measures only the transverse spin component before the second transient. If the first transient also leaves $s_{z,\infty}=0$, only the $J_\nu$ branch remains: $s_y\propto e^{-\gamma_\perp t_1}$ and $s_z\propto e^{-(\gamma_\perp+\kappa/2)t_1}$, so the asymptotic fringe-decay time is $1/\gamma_\perp$. This balanced-population condition for resonant spins does not extend to the entire detected ensemble. Near-resonant spins within the cavity's finite spectral range undergo different rotations due to their different detunings from the cavity field and can retain population imbalance with nonzero $s_{z,\infty}$, allowing the remaining cavity field to generate coherence with a decay timescale set by $\kappa/2$ as derived above. Their net contribution to the readout signal depends on the frequency distribution, rotation angle and axis, and readout weights.

This reduced model illustrates a longer-lived driven response, rather than establishing preservation of arbitrary initial coherence or collective cavity protection. Quantitative envelopes also depend on transient overlap, residual detuned driving, and spectral readout weighting; a Lorentzian frequency distribution is not generally equivalent to homogeneous dephasing during driven evolution. A noise-based treatment should further account for environmental and drive-noise spectra, following the single-drive relaxation analysis in Eq.~(D.5) of Ref.~\cite{wang_coherence_2020}.

\subsection{Phase-jump-induced spin-echo dynamics}
\label{sec:phase_jump_echo}

We next extend the cavity-generated Ramsey sequence by introducing a phase jump of the continuous microwave drive at the midpoint of the evolution. Because the intracavity field cannot respond instantaneously, the phase discontinuity generates an additional cavity transient. For a $\pi$ phase jump, this transient has approximately twice the amplitude of an ordinary ring-up or ring-down transient and therefore naturally provides the intermediate rotation of a nominal $\pi/2$--$\pi$--$\pi/2$ sequence.

We first derive the cavity field for an arbitrary midpoint phase jump. We then isolate the interference generated by the three transients and identify the ideal refocusing limit. Figure~\ref{fig:S1} shows the population and Fourier maps obtained with the full finite-duration cavity response.

\subsubsection{Cavity response to a midpoint phase jump}

We derive the intracavity field throughout the phase-jump sequence. We continue to work in the cavity rotating frame where the cavity amplitude obeys
\begin{equation}
    \dot{\alpha}^{(c)}(t_1) = -\frac{\kappa}{2}\alpha^{(c)}(t_1) + \eta q(t_1)e^{-i\Delta_{\mathrm{dc}}t_1}.
    \label{eq:cavity_eom_phasejump}
\end{equation}
Here $\eta$ is the reference input amplitude and $q(t_1)$ is the dimensionless complex envelope used in the main text. For this sequence, $q=1$ in the first half, $q=e^{i\phi}$ in the second half, and $q=0$ outside the pulse. 

The sequence consists of three stages. The drive is switched on at $t_1=0$, its phase is abruptly changed by an angle $\phi$ at $t_1=\tau=t/2$, and the drive is switched off at $t_1=t=2\tau$. Starting from an empty cavity, $\alpha^{(c)}(0)=0$, the field during the first half is $   \alpha_{1}^{(c)}(t_1) = \bar{\alpha} \left( e^{-i\Delta_{\mathrm{dc}}t_1} - e^{-\kappa t_1/2} \right)$, for $ 0<t_1<\tau$.
At $t_1=\tau$, the drive phase changes from $0$ to $\phi$, so that the new forced solution is $\bar{\alpha} e^{i\phi}e^{-i\Delta_{\mathrm{dc}}t_1}$. Continuity of the intracavity field requires the difference between $\alpha_{1}^{(c)}(\tau)$ and this new forced solution to appear as a new cavity transient. Thus, for $\tau<t_1<t$,
\begin{align}
    \alpha_{2}^{(c)}(t_1) &=\bar{\alpha} e^{i\phi}e^{-i\Delta_{\mathrm{dc}}t_1} + \left[ \alpha_{1}^{(c)}(\tau) - \bar{\alpha} e^{i\phi}e^{-i\Delta_{\mathrm{dc}}\tau} \right] e^{-\kappa(t_1-\tau)/2},\nonumber\\
    &=\bar{\alpha} e^{i\phi}e^{-i\Delta_{\mathrm{dc}}t_1} + \bar{\alpha} \left[ (1-e^{i\phi})e^{-i\Delta_{\mathrm{dc}}\tau} - e^{-\kappa\tau/2} \right] e^{-\kappa(t_1-\tau)/2}.    \label{eq:cavity_second_half}
\end{align}
The complex amplitude of the transient immediately after the phase jump is therefore $\beta_{\rm pj} = \bar{\alpha} \left[ (1-e^{i\phi})e^{-i\Delta_{\mathrm{dc}}\tau} - e^{-\kappa\tau/2} \right]$. When the cavity has nearly reached its driven steady state before the phase jump, $e^{-\kappa\tau/2}\ll1$, this becomes
\begin{equation}
    \beta_{\rm pj} \simeq \bar{\alpha}(1-e^{i\phi})e^{-i\Delta_{\mathrm{dc}}\tau}, \qquad |\beta_{\rm pj}| \simeq 2|\bar{\alpha}| \left|\sin\frac{\phi}{2}\right|.
    \label{eq:phase_jump_amplitude}
\end{equation}
A $\pi$ phase jump therefore produces a transient with approximately twice the amplitude of an ordinary turn-on or turn-off transient. When the latter are adjusted to generate $\pi/2$ rotations, the phase-jump transient correspondingly generates approximately a $\pi$ rotation. The three switching events then realize the nominal sequence $\frac{\pi}{2}-\pi-\frac{\pi}{2}$. Importantly, the intermediate rotation is generated without increasing the microwave amplitude; it results from the mismatch between the field stored in the cavity and the new steady state imposed by the phase-shifted drive.

Immediately before the final switch-off at $t=2\tau$, the cavity field is $\alpha^{(c)}(t^-) =\bar{\alpha}\Big[e^{i\phi}e^{-i2\Delta_{\mathrm{dc}}\tau} +(1-e^{i\phi}) e^{-\kappa\tau/2}e^{-i\Delta_{\mathrm{dc}}\tau}-e^{-\kappa\tau} \Big]$. 
After the drive is switched off, the remaining field freely rings down at the cavity frequency, $\alpha_{3}^{(c)}(t_1)=\alpha^{(c)}(t^-)e^{-\kappa(t_1-t)/2}$ for $t_1>t$. For the spin-echo sequence, $\phi=\pi$, and the cavity field simplifies to $\alpha^{(c)}(t^-) = -\bar{\alpha} \left( e^{-i\Delta_{\mathrm{dc}}\tau} - e^{-\kappa\tau/2} \right)^2$. For well-separated transients, the final ring-down field is $-\bar{\alpha}e^{-i2\Delta_{\mathrm{dc}}\tau}e^{-\kappa t_2/2}$. It has the same envelope and rotation magnitude as the initial transient, with its transverse axis shifted by $-2\Delta_{\mathrm{dc}}\tau=-\Delta_{\mathrm{dc}}t$ in the cavity rotating frame.

\subsubsection{Evolution dynamics}

To isolate the effect of the switching transients, we omit the direct action of the forced field on the spin while retaining the transient waveforms generated by the applied drive. The relative phases of these transients still depend on the drive phase at their launch times. A scan of the total pulse duration therefore changes the interference between the transient-induced rotations even when the forced field does not directly drive the spin.

For well-separated transients and $|\Delta_{\mathrm{sc}}|\ll\kappa$, the detuning term is negligible during each transient. Integrating the envelopes gives signed rotation areas $-\theta,2\theta,-\theta$, where $\theta=2\Omega_d/\kappa$. These rotations are derived in the cavity frame. 
The completed transient operations and the intervening free evolution then give
\begin{equation}
    \mathcal U_{\rm tr}^{(c)}(t)\simeq
    R_{-2\Delta_{\mathrm{dc}}\tau}(-\theta)\,
    R_z(-\Delta_{\mathrm{sc}}\tau)\,
    R_{-\Delta_{\mathrm{dc}}\tau}(2\theta)\,
    R_z(-\Delta_{\mathrm{sc}}\tau)\,
    R_0(-\theta).
    \label{eq:echo_transient_only_propagator}
\end{equation}
All axes and free-evolution operators in this expression are defined in the cavity frame, with $R_z(\beta)=e^{-i\beta\sigma_z/2}$, $R_{\phi}(\beta)$ a transverse rotation about the $\phi$ axis by an angle $\beta$. To evaluate the population, use $R_\phi(\beta)=R_z(\phi)R_x(\beta)R_z(-\phi)$ and combine adjacent $z$ rotations. Defining $\varphi=(\Delta_{\mathrm{dc}}-\Delta_{\mathrm{sc}})\tau$, we obtain
\begin{align}
    \mathcal U_{\rm tr}^{(c)}(t)
    &\simeq R_z(-\Delta_{\mathrm{dc}}t)\,\widetilde U(t),\quad\widetilde U(t)
    =R_x(-\theta)R_z(\varphi)R_x(2\theta)
    R_z(\varphi)R_x(-\theta).
    \label{eq:echo_transient_reduced_propagator}
\end{align}
The leftmost $z$ rotation leaves the final populations unchanged. Multiplying the matrices in $\widetilde U$ for an initial state $|0\rangle$ gives
\begin{align}
    \langle1|\widetilde U(t)|0\rangle
    =i\sin\theta\cos\theta\,(\cos\varphi-1) =-i\sin(2\theta)\sin^2(\varphi/2).
    \label{eq:echo_transient_transition_amplitude}
\end{align}
Taking the absolute square and using $t=2\tau$ then yields
\begin{equation}
    P_{1,\rm tr}(t)=
    \left|\langle1|\mathcal U_{\rm tr}^{(c)}(t)|0\rangle\right|^2
    =\sin^2(2\theta)
    \sin^4\!\left[\frac{(\Delta_{\mathrm{dc}}-\Delta_{\mathrm{sc}})t}{4}\right].
    \label{eq:echo_transient_only_population}
\end{equation}
The difference $\Delta_{\mathrm{dc}}-\Delta_{\mathrm{sc}}=\Delta_{\mathrm{ds}}$ therefore appears in the interference signal while the transient-induced rotations remain near-resonant cavity-frequency operations. Expanding $\sin^4 x=3/8-\cos(2x)/2+\cos(4x)/8$ gives components at angular frequencies $|\Delta_{\mathrm{ds}}|/2$ and $|\Delta_{\mathrm{ds}}|$. This simplified result assumes well-separated transients and small spin--cavity detuning during each rotation.

For $\theta=\pi/2$, the ideal instantaneous sequence gives complete refocusing and $P_{1,\rm tr}=0$. For other rotation angles, the same sequence produces residual population oscillations, with maximal prefactor at $\theta=\pi/4$. In the full cavity response, finite transient duration and overlap at short pulse spacings modify the effective operations. Spins away from the cavity resonance also experience different rotation angles and evolving axes during each transient. These effects allow an oscillatory response even when the nominal drive strength corresponds to $\pi/2$ rotations for an isolated cavity-resonant edge transient.

\begin{figure}
    \centering
    \includegraphics[width=0.95\linewidth]{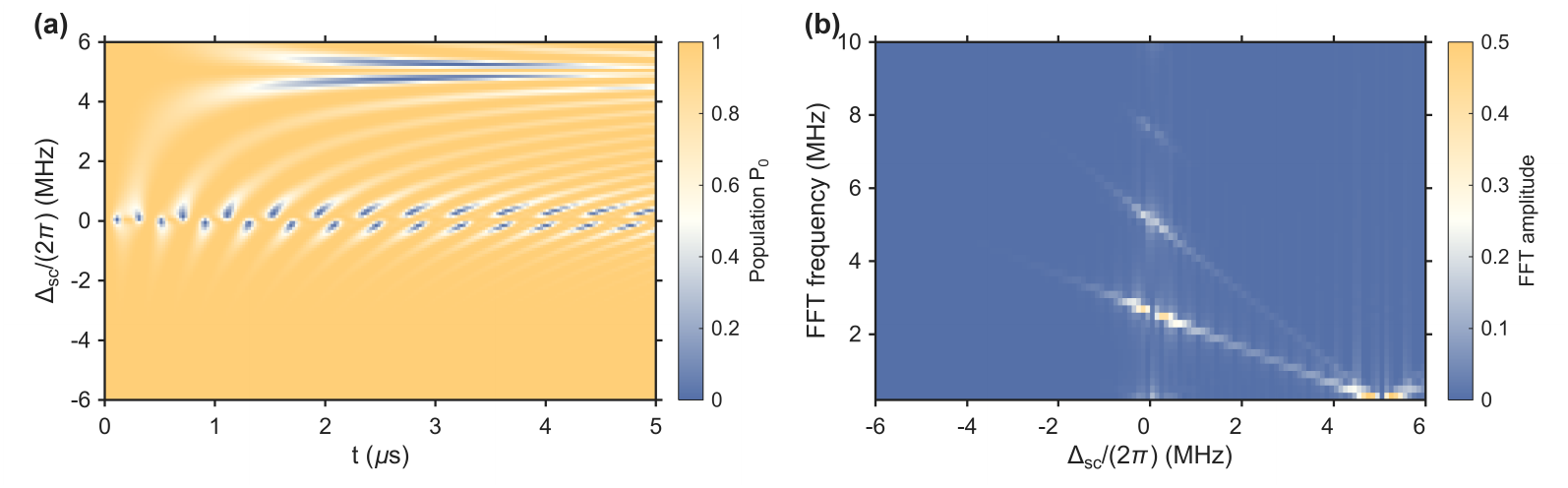}
    \caption{\textbf{Simulated spin population and Fourier spectrum for a midpoint drive-phase reversal.}
    A square microwave pulse of total duration $t$ undergoes a $\pi$ phase jump at $t/2$.
    (a) Population $P_0$ as a function of $t$ and spin--cavity detuning $\Delta_{\mathrm{sc}}=\omega_s-\omega_c$, starting from $|0\rangle$. Each population is evaluated after a fixed delay following drive switch-off, chosen long enough for the final cavity ring-down to finish. The color scale denotes population.
    (b) Single-sided Fourier amplitude of the population traces along the pulse-duration axis at each detuning. The prominent branch occurs near $|\Delta_{\rm ds}|/2$, corresponding to half the drive--spin detuning, with a weaker component near $|\Delta_{\rm ds}|$.
    The simulation includes the forced cavity field and finite switching transients and neglects intrinsic spin relaxation and dephasing. Parameters are $\Delta_{\mathrm{dc}}/(2\pi)=5\,\mathrm{MHz}$, $\kappa/(2\pi)=0.3\,\mathrm{MHz}$, and $\Omega_d=\pi\kappa/4$, giving an isolated cavity-resonant transient area of $\pi/2$.}
    \label{fig:S1}
\end{figure}

Figure~\ref{fig:S1} shows a calculation retaining the full cavity field, including the forced response and all three finite transients. For each spin detuning and applied-pulse duration, the spin is initialized in $|0\rangle$, propagated through the drive and phase reversal, and read out after a fixed post-switch-off delay long enough for the cavity field to decay. Intrinsic spin relaxation and dephasing are neglected. The persistence of oscillations in numerical checks with the forced-field contribution removed is consistent with the transient-only interference mechanism above. In the experiment, the broadened ensemble and the frequency dependence of the microwave readout weight the contributions of the different spin detunings.

\section{Experiment}
\subsection{Experimental platform}
\begin{figure*}[h]
\centering

\begin{tikzpicture}[
    font=\sffamily\small,
    >=Latex,
    inst/.style={
        draw,
        line width=0.8pt,
        rounded corners=1pt,
        minimum width=2.35cm,
        minimum height=1.00cm,
        align=center
    },
    switch/.style={
        draw,
        line width=0.8pt,
        rounded corners=1pt,
        minimum width=1.45cm,
        minimum height=0.80cm,
        align=center
    },
    amp/.style={
        draw,
        isosceles triangle,
        isosceles triangle apex angle=55,
        shape border rotate=90,
        minimum width=1.35cm,
        minimum height=1.65cm,
        inner sep=1pt,
        line width=1.0pt,
        scale=0.8,
        transform shape
    },
    sample/.style={
        draw,
        line width=0.8pt,
        rounded corners=1pt,
        minimum width=2.8cm,
        minimum height=3.2cm,
        align=center
    },
    iq/.style={
        draw,
        line width=0.8pt,
        rounded corners=1pt,
        minimum width=7.2cm,
        minimum height=1.7cm,
        align=center
    },
    signal/.style={
        line width=0.9pt,
        -{Latex[length=2.0mm,width=1.4mm]}
    },
    control/.style={
        dashed,
        line width=0.75pt,
        -{Latex[length=1.8mm,width=1.3mm]}
    },
    textlabel/.style={
        font=\scriptsize,
        fill=white,
        inner sep=1.5pt
    }
]

\node[inst] (lmx) at (0,4.0) {
    \textbf{LMX2572}\\
    Signal Generator
};

\node[amp,shape border rotate=0] (hpamp) at (2.75,4.0) {};

\node[
    font=\scriptsize,
    align=center,
    above=3mm of hpamp
] {
    \textbf{ZHL-16W-43-S+}\\
    High-Power Amplifier
};

\node[switch] (drivesw) at (5.3,4.0) {
    RF Switch
};

\node[inst] (slfsp) at (-0.4,1.7) {
    \textbf{SLFSP06A}\\
    Low-Noise Signal Generator
};

\node[inst,minimum width=1.9cm] (splitter) at (3.0,1.7) {
    RF\\
    Splitter
};

\node[switch] (probesw) at (5.3,1.7) {
    RF Switch
};

\node[
    inst,
    minimum width=1.75cm,
    minimum height=0.90cm
] (combiner) at (8.2,3) {
    RF\\
    Combiner
};

\node[
    circle,
    draw,
    line width=0.8pt,
    minimum size=1.0cm,
    inner sep=0pt
] (circ) at (8.2,1.7) {};

\draw[line width=0.8pt]
    ($(circ.center)+(140:0.30cm)$)
    arc[start angle=140,end angle=-125,radius=0.30cm];

\draw[
    line width=0.8pt,
    -{Latex[length=1.5mm,width=1.15mm]}
]
    ($(circ.center)+(-125:0.30cm)$)
    -- ++(-215:0.08cm);
    
\node[sample] (sample) at (13.0,2.85) {
    \textbf{Microwave Coil}
    \\[7mm]
    \textbf{Dielectric Cavity}
    \\
    +
    \\
    \textbf{NV Ensemble}
};

\node[switch] (rxsw) at (8.2,0.5) {
    RF Switch
};

\node[amp,shape border rotate=270] (rfamp) at (8.2,-0.8) {};

\node[
    font=\scriptsize,
    align=center,
    right=1mm of rfamp
] {
    RF AMP
};

\node[amp,shape border rotate=270] (loamp) at (3.0,-0.8) {};

\node[
    font=\scriptsize,
    align=center,
    right=1mm of loamp
] {
    LO AMP
};

\node[iq] (adl) at (5.6,-3.2) {
    {\large\textbf{ADL5380}}
    \\[1mm]
    IQ Demodulator
};

\draw[signal]
    (lmx.east)
    -- (hpamp.west);

\draw[signal]
    (hpamp.east)
    -- (drivesw.west);

\draw[signal]
    (drivesw.east)
    -- node[above=5mm,textlabel] {Drive}
    ++(0.55,0)
    |- ($(combiner.west)+(0,0.1)$) ;

\draw[signal]
    (slfsp.east)
    -- (splitter.west);

\draw[signal]
    (splitter.east)
    -- (probesw.west);

\draw[signal]
    (probesw.east)
    -- node[below = 5mm,textlabel] {Probe}
    ++(0.55,0)
    |- ($(combiner.west)-(0,0.1)$);

\draw[signal]
    (combiner.south)
    -| (circ.north);

\draw[signal]
    ($(circ.east)+(0,0.17)$)
    --
    node[above=1mm,textlabel] {Drive + Probe}
    ($(sample.west)+(0,-0.99)$);

\draw[signal]
    ($(sample.west)-(0,1.32)$)
    --
    node[below=1mm,textlabel] {Cavity reflection}
    ($(circ.east)+(0,-0.17)$);

\draw[signal]
    (circ.south)
    -- (rxsw.north);

\draw[signal]
    (rxsw.south)
    -- (rfamp.north);

\draw[signal]
    (splitter.south)
    -- (loamp.north);

\draw[signal]
    (loamp.south)
    -- node[left,textlabel] {LO}
    ($(adl.north)+(-2.6,0)$);

\draw[signal]
    (rfamp.south)
    -- node[right,textlabel] {RF}
    ($(adl.north)+(2.6,0)$);

\node[
    font=\bfseries\small,
    anchor=west
] at (-1,4.9) {
    Spin drive
};

\node[
    font=\bfseries\small,
    anchor=west
] at (-2.5,2.55) {
    Probe and cavity readout
};

\node[
    font=\bfseries\small,
    anchor=west
] at (-0.5,-3) {
    IQ detection
};

\end{tikzpicture}

\caption{\textbf{Microwave generation, cavity readout, and IQ detection.}
During control, microwave pulses generated by an LMX2572 are amplified and gated by an RF switch before reaching the cavity through a circulator. During readout, a weak probe from an SLFSP06A is sent to the cavity through the same path, while a second branch serves as the local oscillator (LO). The reflected probe passes through a protection switch and an RF amplifier before entering the ADL5380 IQ demodulator. Its differential $I$ and $Q$ outputs are digitized by an NI USB-6451 DAQ (not shown).
}
\label{fig:S2}
\end{figure*}
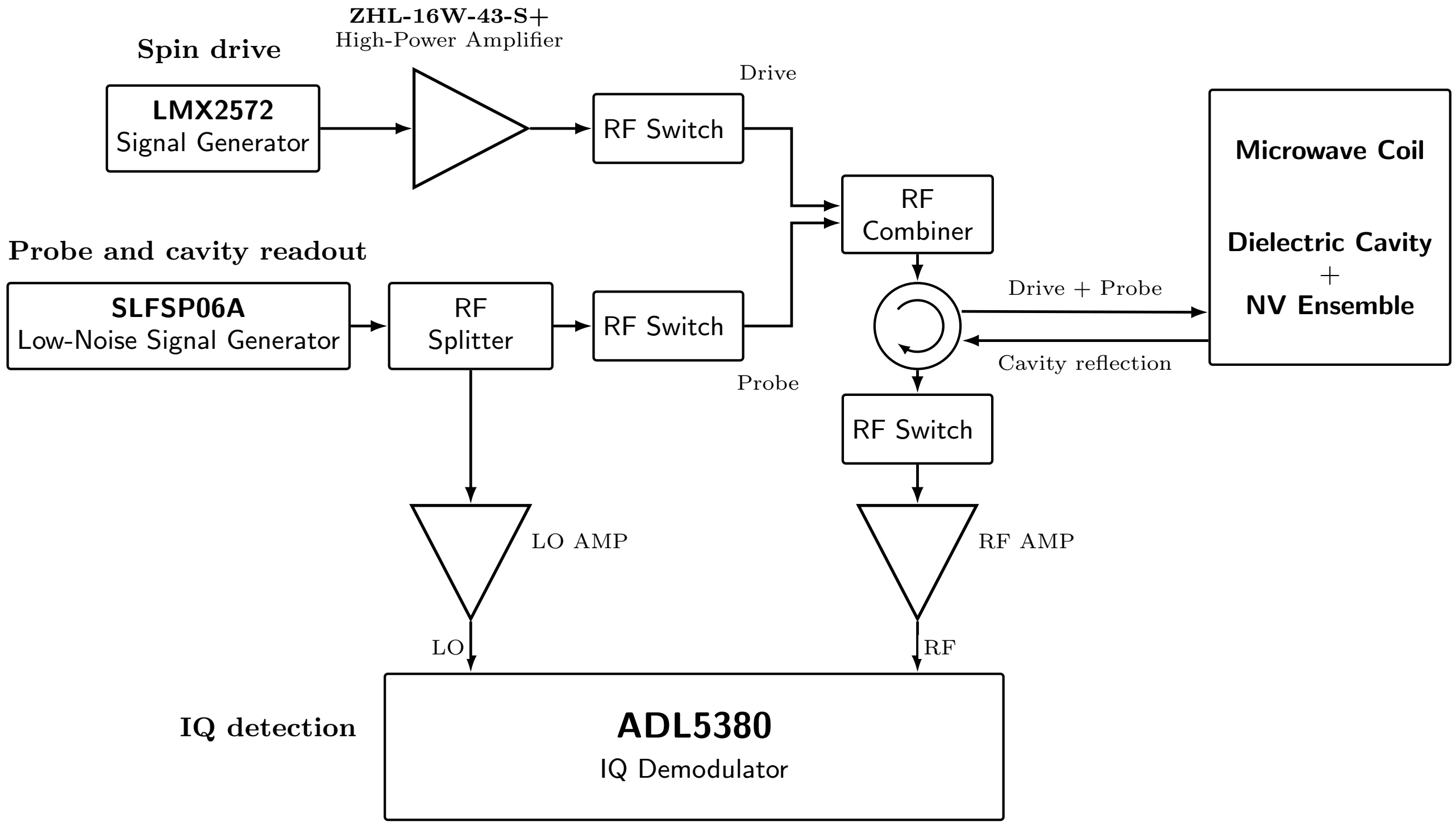

Our experimental platform consists of an NV ensemble in diamond coupled to a dielectric resonator. A 532~nm laser with a power of approximately 5W optically polarizes the ensemble toward the $\lvert m_s=0\rangle$ ground state. A strong microwave drive and a weak microwave probe share the loop antenna coupled to the resonator. The reflection of microwave probe is measured using IQ demodulation, with a second branch of the probe signal serving as the local oscillator. A pair of permanent magnets (NdFeB magnet, $50~\mathrm{mm}$ in diameter and $5~\mathrm{mm}$ thick) provide a static bias field that tunes the NV transitions close to the cavity resonance, while tuning coils provide fine adjustment of the spin--cavity detuning over a limited range. The microwave generation and detection circuit is shown in Fig.~\ref{fig:S2}.

The hollow cylindrical MgO--TiO$_2$ dielectric resonator has an axial height of $8.175~\mathrm{mm}$, an outer diameter of $16.5~\mathrm{mm}$, and a bore diameter of $6.2~\mathrm{mm}$. The measured resonance frequency is $2.9868~\mathrm{GHz}$, with a loaded cavity linewidth of approximately $250~\mathrm{kHz}$. The sample is a (100)-oriented single-crystal CVD diamond (Element Six, DNV-B14), with nominal dimensions of $3\times3\times0.5~\mathrm{mm^3}$ and a typical NV concentration of $4~\mathrm{ppm}$. The diamond is positioned at the center of the resonator bore, as shown in Fig.~1(a).

\subsection{IQ detection and data acquisition}

\textit{Cavity reflection measurement.} The output capacitors of the ADL5380 quadrature demodulator were removed to allow DC coupling. The outputs $I_{\mathrm H}, I_{\mathrm L}$ and $Q_{\mathrm H}, Q_{\mathrm L}$ were connected directly to two differential channels of an NI USB-6451 DAQ, which simultaneously recorded $I(t)=V_{I_{\mathrm H}}-V_{I_{\mathrm L}}$ and $Q(t)=V_{Q_{\mathrm H}}-V_{Q_{\mathrm L}}$.
The complex signal was defined as $Z(t)=I(t)+iQ(t)$, with amplitude $|Z|$ and phase $\phi=\arg Z$. Because the local oscillator and cavity probe were derived from the same microwave source, the demodulated signal represents the complex cavity response in the probe rotating frame.

\textit{Receiver protection and recovery.} During high-power spin manipulation, the receiver path was isolated by an RF protection switch in order to prevent saturation of the receive amplifier and the ADL5380. The drive-path and receiver-path switches were controlled synchronously by the pulse programmer. During the high-power microwave pulse the drive path was enabled while the receiver path was blocked; after the drive pulse the drive path was disabled and the receiver path was restored.

\textit{Microwave readout signal extraction.} The spin ensemble modifies the cavity response to the weak microwave probe. The modification is dependent on the spin populations prepared by the control sequence. The reflected field is related to the intracavity field through the input--output relation, $a_{\mathrm{out}}=a_{\mathrm{in}}-\sqrt{\kappa_{\mathrm{ext}}}\alpha$. Unless otherwise specified, the microwave readout signal is obtained by averaging the measured quadrature signal $Q(t)$ over the readout window and subtracting its value at the end of the window, $S_Q=\frac{1}{t_2-t_1}\int_{t_1}^{t_2}Q(t)\,\mathrm{d}t-Q(t_2)$, where $t_1=20~\mu\mathrm{s}$ and $t_2=560~\mu\mathrm{s}$. This procedure provides spectral selectivity, as discussed in Sec.~\ref{sec:sim_readout}. The resulting signal is further normalized by a reference value corresponding to $|r|\simeq 1$.

\subsection{Experimental Calibrations}
\subsubsection{Magnetic-field alignment by continuous-wave spectroscopy}

\begin{figure}[h]
    \centering
    \includegraphics[width=0.5\textwidth]{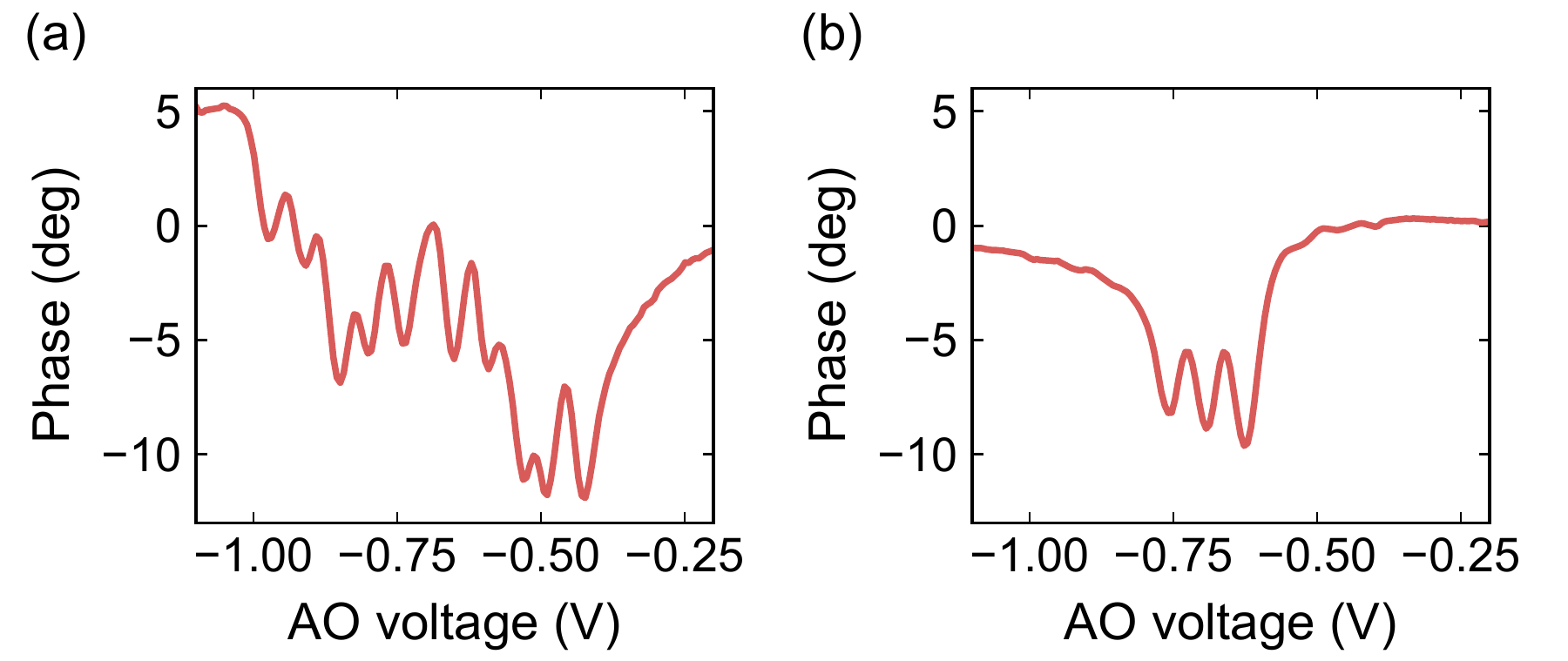}
    \caption{\textbf{Magnetic field alignment using continuous-wave spin spectroscopy.}
    (a) Before alignment, the four NV classes exhibit multiple separated resonance groups.
    (b) After alignment, the corresponding resonance peaks merge into a single set of hyperfine-resolved resonances, indicating spectroscopic degeneracy of the NV classes. 
    }
    \label{fig:S3}
\end{figure}

The NV ensemble contains four crystallographic orientations along the diamond $\langle111\rangle$ axes. The magnetic field was aligned to give equal longitudinal projection magnitudes along the four NV orientations, bringing their spin resonances into overlap. This was achieved by adjusting two angles: rotation about the cavity $z$ axis and rotation of the magnet pair about an axis perpendicular to it. The alignment was checked using continuous-wave spectroscopy. As shown in Fig.~\ref{fig:S3}, the initially separated resonances (a) merge into a single set of hyperfine-resolved peaks after alignment (b).

\subsubsection{Cavity characterization}

The bare cavity response was characterized using a weak microwave probe with the optical excitation disabled and the NV ensemble magnetically detuned from the cavity resonance, where the spin-induced modification of the cavity susceptibility was negligible. The microwave probe power was kept in the linear-response regime throughout the calibration.

The measured cavity response was fitted to the susceptibility $\chi_c(f)  = \frac{1}{\kappa/2-i2\pi(f-f_c)}$, together with a slowly varying complex background accounting for the frequency response and electrical phase of the microwave chain. The fit yields the bare cavity resonance frequency $f_c$ and the loaded cavity decay rate $\kappa$.

\begin{figure}[h]
    \centering
    \includegraphics[width=1\textwidth]{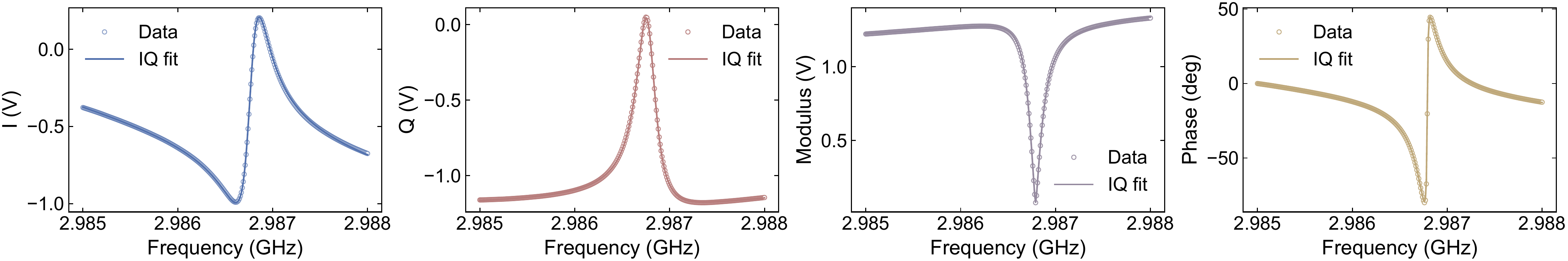}
    \caption{\textbf{Cavity characterization and fit}. The cavity response was measured using a weak microwave probe with the optical excitation disabled and the NV ensemble detuned from the cavity resonance. The solid curve denotes the fit to the cavity response model.}
    \label{fig:S4}
\end{figure}

\subsubsection{Spin spectroscopy and frequency calibration}
The NV transition frequencies were fine-tuned by varying $V_{\mathrm{AO}}$, which controlled the current in the magnetic-field tuning coils via a voltage-to-current converter. To calibrate the NV spin-transition frequencies, we acquired an ODMR spectrum at each voltage with the spins detuned from the cavity and extracted the center frequencies of the three resolved hyperfine resonances.

The calibration measurements were performed using the same optical duty cycle and sequence dead time as the main measurements. This choice reproduces, to first order, the steady-state thermal conditions established by laser illumination and sequence repetition, thereby incorporating the corresponding thermally induced shift of the NV resonance into the frequency calibration. The spin resonance was additionally kept sufficiently detuned from the cavity during the calibration. In this regime, a lower microwave power was sufficient for ESR detection, reducing power broadening and allowing the hyperfine components to be resolved more accurately.

For each hyperfine branch, the resonance frequency was found to vary approximately linearly with the magnetic-field control voltage, and was parameterized as
\begin{equation}
    f_{s,m_I}(V_{\mathrm{AO}})
    =
    a_{m_I}V_{\mathrm{AO}} + b_{m_I},
    \qquad m_I=-1,0,+1,
\end{equation}
where $a_{m_I}$ and $b_{m_I}$ are obtained from independent linear fits to the three hyperfine branches. For the central component ($m_I=0$), the fit yields
\begin{equation}
    f_{s,0}(V_{\mathrm{AO}})
    =
    2999.35~\mathrm{MHz}
    -
    (18.32~\mathrm{MHz/V})V_{\mathrm{AO}}.
    \label{eq:central-hyperfine-calibration}
\end{equation}

These relations were subsequently used to convert the programmed field-control voltage into the corresponding spin-transition frequency and, where required, the cavity--spin detuning
\begin{equation}
    \Delta_{\mathrm{sc}}
    =
    2\pi\left(f_{s,0}-f_c\right).
\end{equation}
for consistency, we define the spin frequency $f_s$ entering the detuning as the resonance frequency of the central hyperfine component ($m_I=0$).

\begin{figure}[t]
    \centering
    \includegraphics[width=0.6\textwidth]{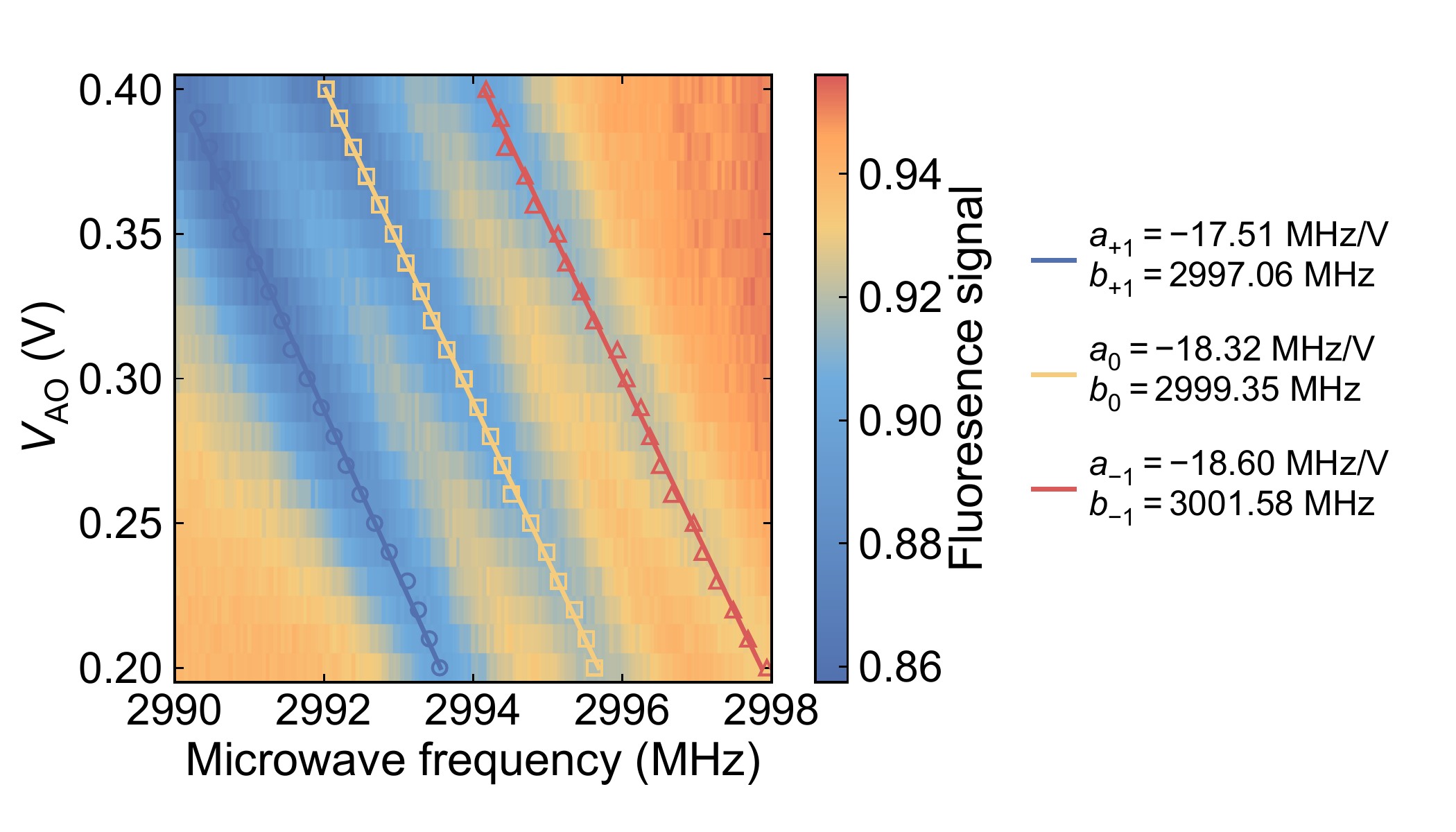}
    \caption{\textbf{Spin spectroscopy and magnetic-field frequency calibration.} Extracted resonance frequencies of the three hyperfine branches as a function of the magnetic-field control voltage $V_{\mathrm{AO}}$. Solid lines denote linear fits used to convert the experimental control voltage into spin-transition frequency.
    }
    \label{fig:S5}
\end{figure}

\subsubsection{Microwave amplitude and pulse-strength calibration}

To calibrate the effective microwave driving strength, we measured the output power of the microwave signal source at different settings and converted each power measurement to a corresponding microwave amplitude. The Lorentzian cavity response imposes frequency-dependent filtering on the applied microwave drive. Neglecting spin backaction, we obtain the steady-state intracavity field $\bar{\alpha}$ from the long-time limit of Eq.~\eqref{eq:alpha_ringup}, and the effective driving strength is proportional to the local microwave magnetic-field amplitude and hence to $\lvert\bar{\alpha}\rvert$.

For negligible spin-drive detuning and an approximately constant input drive amplitude over the calibration frequency range, the measured Rabi frequency satisfies
\begin{equation}
    f_R(\Delta_{\mathrm{dc}})
    =
    \frac{\Omega_d(\Delta_{\mathrm{dc}})}{2\pi}
    =
    \frac{\Omega_0}{2\pi}
    \frac{\kappa/2}{\sqrt{\Delta_{\mathrm{dc}}^{\,2}+(\kappa/2)^2}}.
    \label{eq:Rabi_calibration}
\end{equation}
Here, $\Omega_0$ denotes the driving strength on cavity resonance, whereas $\Omega_d$ denotes the steady-state driving strength at the actual drive--cavity detuning, the loaded cavity linewidth is $\kappa/(2\pi)=0.25~\mathrm{MHz}$. 
We measured Rabi oscillations at different drive--cavity detunings while maintaining a fixed microwave power (reaching the cavity) of $37.8~\mathrm{dBm}$. A nonlinear least-squares fit to the measured Rabi frequencies as a function of detuning yielded an on-resonance Rabi frequency of $\Omega_0/(2\pi)=21.4~\mathrm{MHz}$. With the microwave transmission and cavity coupling conditions unchanged, the driving strength $\Omega_d$  at other microwave drive power and detuning conditions is therefore obtained using the above calibration as a reference. 
This calibration relates the $\Omega_0$ and $\Omega_d$, providing an experimental amplitude scale for the cavity-transient spin dynamics.

\section{Simulation}
\subsection{Numerical simulation of the Ramsey-to-Rabi crossover}
\label{sec:sim_dynamics}
We implement numerical simulations based on the model developed in Sec.~\ref{sec:ramsey_limit}. The intracavity microwave field is calculated using the analytical response of an effectively empty cavity to a rectangular driving pulse, as described by Eqs.~\eqref{eq:alpha_ringup} and \eqref{eq:alpha_ringdown_exact}. The resulting field drives an ensemble of independent effective two-level spin systems. Spin--spin interactions, spin relaxation and spin decoherence are neglected.
Each spin is initialized in $|0\rangle$ and propagated numerically using the midpoint value of the evolution operator,
$|\psi(t'+\delta t')\rangle=\exp\left[-iH\left(t'+\frac{\delta t'}{2}\right)\delta t'\right]|\psi(t')\rangle$, where $H(t')$ is denoted in Eqs.~\eqref{eq:H_ringup_vector} and \eqref{eq:H_ringdown_vector}.
The microwave pulse duration $t$ is varied from $0$ to $4\,\mu\mathrm{s}$, and the population $P_0=|\langle 0|\psi(t_{\mathrm{read}})\rangle|^2$ is evaluated at $t_{\mathrm{read}}=20\,\mu\mathrm{s}$, including the cavity ringdown after switch-off.

The simulation parameters are as follows. The cavity resonance frequency and linewidth are $\omega_c/(2\pi)=2986.8\,\mathrm{MHz}$ and $\kappa/(2\pi)=0.25\,\mathrm{MHz}$, respectively. The drive frequency is fixed at $\omega_d/(2\pi)=2993.8\,\mathrm{MHz}$, giving a drive--cavity detuning of $7\,\mathrm{MHz}$. We use $\Omega_0/(2\pi)=21.4\,\mathrm{MHz}$ for the Rabi frequency on cavity resonance. The actual Rabi frequency is obtained by multiplying $\Omega_0$ by the normalized complex cavity response. The numerical time step is $\delta t'=5\,\mathrm{ns}$.

The NV ensemble is modeled by three equally weighted hyperfine transitions separated by $A_{zz}/(2\pi)=2.16~\mathrm{MHz}$, each with a Lorentzian linewidth of $\Gamma/(2\pi)=1.385~\mathrm{MHz}$ (FWHM). The spin frequency distribution is $\rho(\omega)=\frac{1}{3}\sum_{m=-1}^{1} \frac{\Gamma/(2\pi)}{(\omega-\omega_{s,0}-mA_{zz})^2+(\Gamma/2)^2}$, where $\omega_{s,0}$ is the central hyperfine transition frequency. We sample the distribution on a uniform frequency grid $\omega_j$ and use the weights $W_j=\rho(\omega_j)/\sum_k\rho(\omega_k)$ for ensemble averaging.

\subsection{Dynamical Selectivity in Cavity Readout}

\label{sec:sim_readout}
As discussed in the main text, we can spectrally select what spins to read out by taking advantage of the spin dynamics induced by the weak microwave probe. Spins at different detunings from the cavity resonance experience different effective driving strengths and consequently exhibit different population decay timescales. 
This dynamical evolution provides a mechanism for spectrally selective readout of spins via microwave, as spectral components of spin ensembles contribute differently at different readout time scales.

\begin{figure}[h]
    \centering
    \includegraphics[width=0.98\textwidth]{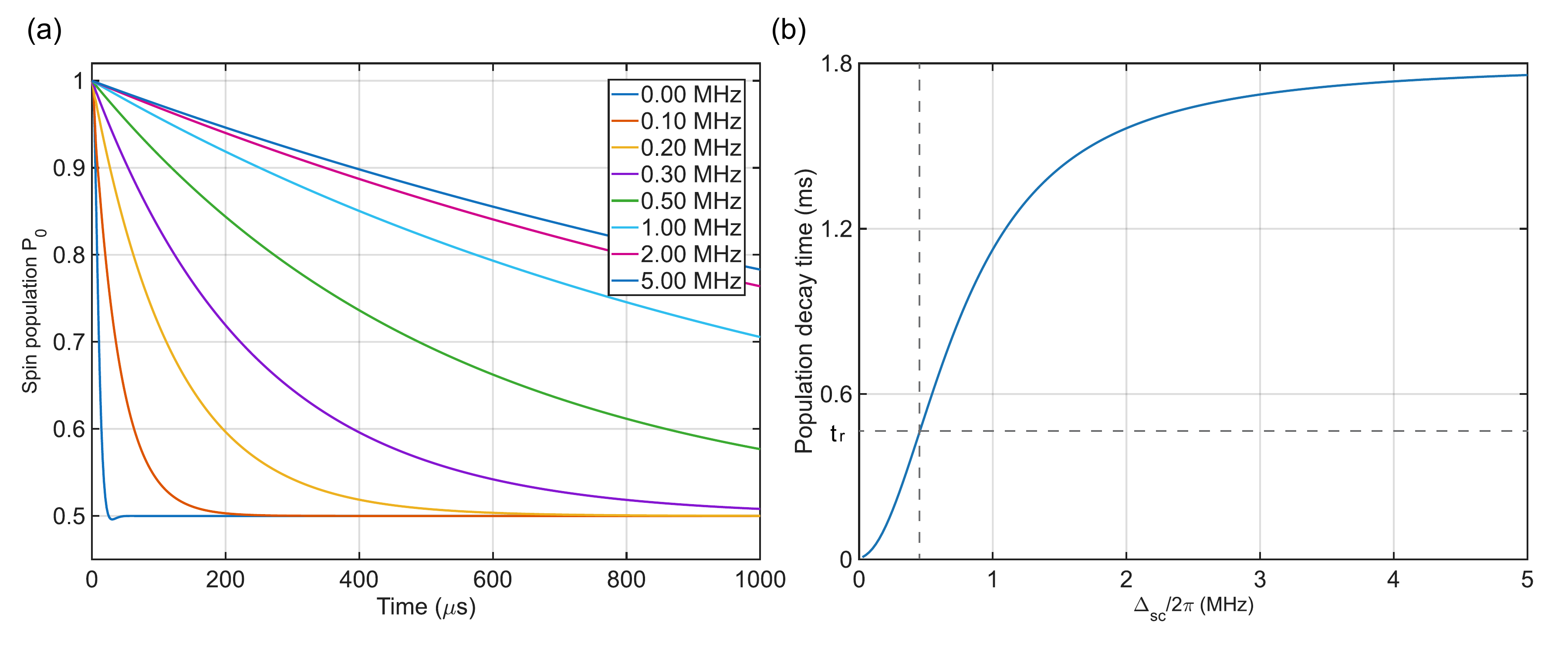}
    \caption{\textbf{Dynamical selectivity in cavity readout.} 
    (a) Simulated evolution of spin population under cavity-resonant microwave probing for different spin detunings.
    (b) Characteristic population decay time as a function of spin detuning, extracted from exponential fits to simulated traces.
    The simulations use $T_1=1800\,\mu\mathrm{s}$, $T_\phi=3\,\mu\mathrm{s}$, $\Omega_p/(2\pi)=0.0317\,\mathrm{MHz}$, and $\kappa/(2\pi)=0.25\,\mathrm{MHz}$.}
    \label{fig:S6}

\end{figure}

We perform numerical simulations using the Lindblad master equation, incorporating spin relaxation and pure dephasing, to investigate the evolution of spin population under a microwave drive resonant with the cavity. For each spin transition, the density matrix evolves according to
\begin{equation}
\frac{d\rho}{dt}
=-\frac{i}{\hbar}[H(t),\rho]
+\frac{\gamma_1}{2}\mathcal{D}[\sigma_-]\rho
+\frac{\gamma_1}{2}\mathcal{D}[\sigma_+]\rho
+\frac{\gamma_\phi}{2}\mathcal{D}[\sigma_z]\rho,
\end{equation}
where $\mathcal{D}[L]\rho=L\rho L^\dagger-\frac{1}{2}\{L^\dagger L,\rho\}$ is the Lindblad dissipator, $\gamma_1=1/T_1$ is the longitudinal relaxation rate, and $\gamma_\phi=1/T_\phi$ is the pure dephasing rate. In the cavity rotating frame, the Hamiltonian is $H(t)=\frac{\hbar}{2}\left[\Delta_{\mathrm{sc}}\sigma_z+\Omega(t)\sigma_x\right]$,
where $\Delta_{\mathrm{sc}}=\omega_s-\omega_c$ is the spin detuning and $\Omega(t)=\Omega_p[1-\exp(-\kappa t/2)]$ accounts for the cavity field ring-up.

As shown in Fig.~\ref{fig:S6}(a), spins with transition frequencies closer to the cavity resonance experience stronger effective driving and consequently exhibit a more rapid decay of population. In contrast, far-detuned spins are only weakly affected by the probe field, and their population decay is predominantly governed by intrinsic relaxation processes, approaching the behavior expected in the absence of driving. To further quantify this detuning dependence, we perform a finer sweep of the spin detuning and extract the characteristic population decay time from exponential fits to the simulated traces. The resulting decay times are plotted as a function of detuning in Fig.~\ref{fig:S6}(b).

This dependence of spin decay on detuning provides an intuitive picture to understand the dynamical selectivity in microwave readout. At a given observation time $t_r$, spins with characteristic decay times longer than $t_r$ retain a larger fraction of their initial population, whereas those with shorter times have already undergone substantial decay. Hence, the surviving population is therefore predominantly associated with spins on the right-hand side of the detuning threshold corresponding to $t_r$ in Fig.~\ref{fig:S6} (b). Furthermore, averaging the measured signal over $[0, t_r]$ and subtracting its value at $t_r$ preferentially selects spins that undergo substantial decay during the observation window, approximately corresponding to those on the left-hand side of the threshold. This picture illustrates how the readout duration can effectively select different spectral components of the spin ensemble.

\subsection{Numerical reproduction of the microwave readout signal}
\label{sec:sim_reproduction}

\begin{figure}[h]
    \centering
\includegraphics[width=\textwidth]{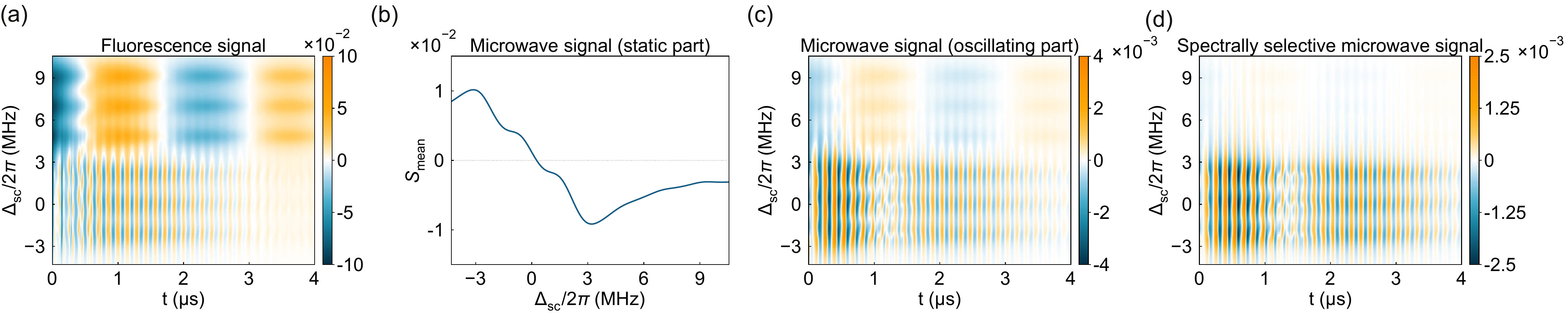}
    \caption{\textbf{Numerical reproduction of the signals in Fig.~2.}
    (a) Simulated fluorescence signal as a function of $t$ and $\Delta_{sc}$, representing the total spin population. 
    (b,c) Background (b) and oscillating (c) components of the simulated microwave reflection signal. Without endpoint subtraction, the oscillating component contains contributions from both Ramsey interference and Rabi oscillations of the driven spins.
    (d) Simulated microwave signal after endpoint subtraction, which preferentially emphasizes spins near the cavity resonance. The resulting signal is dominated by Ramsey interference, with the Rabi features largely suppressed. 
    The heatmaps in (a), (c), and (d) are mean-centered by subtracting the mean of each row.}
    \label{fig:S7}

\end{figure}

We further incorporate the spin dynamics during microwave readout into the Ramsey-to-Rabi crossover simulation. The simulated cavity reflection signal is processed in the same way as the experimental data in Fig.~2. We average $\mathrm{Im}(r)$ over the interval $[20,\upmu\mathrm{s},560,\upmu\mathrm{s}]$ and subtract its value at the end of the readout window. This procedure emphasizes spins closer to the cavity resonance, whose populations change more rapidly during readout. The resulting signal, shown in Fig.~\ref{fig:S7}(d), reproduces the main features of the experimental interference pattern.

These simulations show how the cavity transients and the spin dynamics during microwave readout together shape the observed interference signals. The agreement between simulation and experiment further supports the distinct roles of cavity transients and continuous microwave driving, as well as the contribution of dynamical readout selectivity to the measured signal.

\clearpage

\end{document}